\documentclass[pdflatex,sn-mathphys-num]{sn-jnl}

\usepackage{graphicx}%
\usepackage{multirow}%
\usepackage{amsmath,amssymb,amsfonts}%
\usepackage{amsthm}%
\usepackage{mathrsfs}%
\usepackage[title]{appendix}%
\usepackage{xcolor}%
\definecolor{mediumblue}{rgb}{0.0, 0.0, 0.8}
\usepackage{textcomp}%
\usepackage{manyfoot}%
\usepackage{booktabs}%
\usepackage{algorithm}%
\usepackage{algorithmicx}%
\usepackage{algpseudocode}%
\usepackage{listings}%

\usepackage[braket]{qcircuit}

\usepackage{tikz}
\usetikzlibrary{cd,calc,arrows,shapes.geometric,intersections}
\usetikzlibrary{positioning} 
\usetikzlibrary{decorations.pathmorphing,decorations.markings}
\usetikzlibrary{patterns}
\usetikzlibrary{snakes} 

\usepackage[frozencache=true]{minted} 
\setminted{escapeinside=77}
\usepackage{inconsolata}

\input{mintedblock.tex}

\newcommand\coqin[1]{\text{\mintinline{ssr}{#1}}}

\newcommand\coq{\textsc{Rocq}}
\newcommand\mathcomp{\textsc{MathComp}}

\newcommand\Nat{\mathbb{N}}

\newcommand\finmapsto\mapsto

\newcommand\dpower[2]{{#1}^{\widehat{#2}}}
\newcommand\lens[2]{\texttt{lens}_{#1,#2}}
\newcommand\tuple[2]{{{#1}^{#2}}}
\newcommand\blb{[\hspace{-2.4px}[\hspace{-2.4px}[}
\newcommand\brb{]\hspace{-2.4px}]\hspace{-2.4px}]}
\newcommand\ord[1]{{\blb{#1}\brb}}
\newcommand\lensC[1]{{#1^{\scriptscriptstyle\complement}}}
\newcommand\extractC{\lensC{\texttt{extract}}}
\newcommand\lindex[1]{\texttt{lens\_index}\ {#1}}
\newcommand\tnth[2]{{#1[#2]}}
\let\lcomp\odot
\let\vcomp\bullet
\newcommand\morlin[2]{\texttt{morlin}_{#1,#2}}
\newcommand\mor[2]{\texttt{mor}_{#1,#2}}
\newcommand\ndo[1]{\texttt{endo}_{#1}}

\newcommand\ghzfive{
\Qcircuit @C=1.5em @R=.3em {
 \ket 0 & & \gate H & \ctrl{1} & \qw      & \qw      & \qw      & \qw \\
 \ket 0 & & \qw     & \targ    & \ctrl{1} & \qw      & \qw      & \qw \\
 \ket 0 & & \qw     & \qw      & \targ    & \ctrl{1} & \qw      & \qw \\
 \ket 0 & & \qw     & \qw      & \qw      & \targ    & \ctrl{1} & \qw \\
 \ket 0 & & \qw     & \qw      & \qw      & \qw      & \targ    & \qw
}
}

\newcommand\bitflipcode{
    \Qcircuit @C=0.7em @R=.3em {
       & \ctrl{1} & \ctrl{2} & \multigate{2}{Ch_3} & \ctrl{1} & \ctrl{2} & \targ    & \qw\\
       & \targ    & \qw      & \ghost{Ch_3}        & \targ    & \qw      & \ctrl{-1} & \qw\\
       & \qw      & \targ    & \ghost{Ch_3}        & \qw      & \targ    & \ctrl{-2} & \qw
    }
}

\newcommand\signflipcode{
    \Qcircuit @C=0.7em @R=.3em {
       & \ctrl{1} & \ctrl{2} & \gate{H} & \multigate{2}{Ch_3} & \gate{H} & \ctrl{1} & \ctrl{2} & \targ    & \qw\\
       & \targ    & \qw      & \gate{H} & \ghost{Ch_3}        & \gate{H} & \targ    & \qw      & \ctrl{-1} & \qw\\
       & \qw      & \targ    & \gate{H} & \ghost{Ch_3}        & \gate{H} & \qw      & \targ    & \ctrl{-2} & \qw
    }
}

\newcommand\shorbody{
 & \ket \psi
 &
 & \ctrl{3}
 & \ctrl{6} 
 & \gate{H}
 & \ctrl{1}
 & \ctrl{2}
 & \multigate{8}{Ch_9}
 & \ctrl{1}
 & \ctrl{2}
 & \targ
 & \gate{H}
 & \ctrl{3}
 & \ctrl{6} 
 & \targ
 & \qw\\
 & \ket 0
 &
 & \qw
 & \qw
 & \qw
 & \targ
 & \qw
 & \ghost{Ch_9}
 & \targ
 & \qw
 & \ctrl{-1}
 & \qw
 & \qw
 & \qw
 & \qw
 & \qw\\
 & \ket 0
 &
 & \qw
 & \qw
 & \qw
 & \qw
 & \targ
 & \ghost{Ch_9}
 & \qw
 & \targ
 & \ctrl{-2}
 & \qw
 & \qw
 & \qw
 & \qw
 & \qw\\
 & \ket 0
 &
 & \targ
 & \qw
 & \gate{H}
 & \ctrl{1}
 & \ctrl{2}
 & \ghost{ch}
 & \ctrl{1}
 & \ctrl{2}
 & \targ
 & \gate{H}
 & \targ
 & \qw
 & \ctrl{-3}
 & \qw\\
 & \ket 0
 &
 & \qw
 & \qw
 & \qw
 & \targ
 & \qw
 & \ghost{Ch_9}
 & \targ
 & \qw
 & \ctrl{-1}
 & \qw
 & \qw
 & \qw
 & \qw
 & \qw\\
 & \ket 0
 &
 & \qw
 & \qw
 & \qw
 & \qw
 & \targ
 & \ghost{Ch_9}
 & \qw
 & \targ
 & \ctrl{-2}
 & \qw
 & \qw
 & \qw
 & \qw
 & \qw\\
 & \ket 0
 &
 & \qw
 & \targ
 & \gate{H}
 & \ctrl{1}
 & \ctrl{2}
 & \ghost{Ch_9}
 & \ctrl{1}
 & \ctrl{2}
 & \targ
 & \gate{H}
 & \qw
 & \targ
 & \ctrl{-6}
 & \qw\\
 & \ket 0
 &
 & \qw
 & \qw
 & \qw
 & \targ
 & \qw
 & \ghost{Ch_9}
 & \targ
 & \qw
 & \ctrl{-1}
 & \qw
 & \qw
 & \qw
 & \qw
 & \qw\\
 & \ket 0
 &
 & \qw
 & \qw
 & \qw
 & \qw
 & \targ
 & \ghost{Ch_9}
 & \qw
 & \targ
 & \ctrl{-2}
 & \qw
 & \qw
 & \qw
 & \qw
 & \qw
}

\newcommand\shorcode[1]{
\Qcircuit @C=0.8em @R=.4em {
\shorbody
{#1}
}
}

\newcommand\revcircuitfive{
\Qcircuit @C=2.5em @R=1.1em {
 \ket {\varphi_1} & & \qswap      & \qw         & \qw & \ket {\varphi_5}\\
 \ket {\varphi_2} & & \qw \qwx    & \qswap      & \qw & \ket {\varphi_4}\\
 \ket {\varphi_3} & & \qw \qwx    & \qw \qwx    & \qw & \ket {\varphi_3}\\
 \ket {\varphi_4} & & \qw \qwx    & \qswap \qwx & \qw & \ket {\varphi_2}\\
 \ket {\varphi_5} & & \qswap \qwx & \qw         & \qw & \ket {\varphi_1}
}
}

\theoremstyle{thmstyleone}%
\theoremstyle{thmstyletwo}%

\theoremstyle{thmstylethree}%

\begin{document}

\title[Typed compositional quantum computation with lenses]{Typed compositional quantum computation with lenses}


\author*[1]{\fnm{Jacques} \sur{Garrigue}}\email{garrigue@math.nagoya-u.ac.jp}

\author[1]{\fnm{Takafumi} \sur{Saikawa}}\email{tscompor@gmail.com}
\equalcont{These authors contributed equally to this work.}

\affil*[1]{\orgdiv{Graduate School of Mathematics}, \orgname{Nagoya University}, \orgaddress{\street{Furocho}, \city{Nagoya}, \postcode{464-8602}, \state{Aichi}, \country{Japan}}}


\abstract{
We propose a type-theoretic framework for describing and proving properties
of quantum computations, in particular those presented as quantum circuits.
Our proposal is based on an observation that,
in the polymorphic type system of \coq{}, currying on quantum states
allows one to apply quantum gates directly inside a complex circuit.
By introducing a discrete notion of lens to control this currying,
we are further able to separate the combinatorics of the circuit structure
from the computational content of gates.
We apply our development to define quantum circuits recursively from the
bottom up, and prove their correctness compositionally.
}

\keywords{quantum programming, semantics, lens, currying, Rocq, MathComp}



\maketitle

\section{Introduction\protect\footnotemark}
\footnotetext[\thefootnote]{This article is an extended version of the
presentation with the same title at ITP 2024.}
Quantum computation is a theory of computation whose unit of information is
the state of a quantum particle, called a quantum bit.
A quantum bit is unlike a classical bit in that the former
may retain many values at the same time, albeit they ultimately can
only be observed as probabilities, while the latter has a single value.
This possibility of a multitude of values is preserved
by pure quantum computation, and destroyed by a measurement of the probability.

These properties of quantum bits and computation
are commonly modelled in terms of unitary transformations
in a Hilbert space~\cite{Ying-book2016}.  Such a transformation
is constructed by composing both sequentially and parallelly
various simple transformations called quantum gates.

Many works have been built to allow proving quantum
algorithms in such settings~\cite{Qwire-popl2017,Unruh-planqc21,CoqQ-popl2023},
or more abstractly using string diagrams representing computations in
a symmetric monoidal category~\cite{Coecke-Kissinger-2017}.
We investigate whether some type-theoretic insights could help in
describing and proving properties of quantum computations, in
particular those denoted by so-called quantum circuits.

Our main goal is to reach {\em compositionality} inside a semantical
representation of computations. We wish it both at the level of
definitions and proofs, with as little overhead as possible.
\begin{description}
\item[{\it Definitional compositionality}] means that it should be
  possible to turn any (pure) quantum circuit into an abstract component,
  which can be instantiated repeatedly in various larger circuits.
\item[{\it Proof compositionality}] means that the proof of functional
  properties about (pure) quantum circuits should be statable as a
  generic lemma about the corresponding abstract component,
  so that one can build proofs of a large circuit by applying this lemma
  to instances of the component, without having to unfold the
  concrete definition of the component during the proof.
\item[{\it Abstraction overhead}]
  refers to the extra steps required for abstraction and
  instantiation, both in definitions and proofs.
\end{description}
The approach we have designed represents
circuits as linear transformations, and reaches the above goals by
cleanly separating the complex linear algebra in computation from the
combinatorics of the wiring, using a combinatorial notion of lens.
Compared to more abstract approaches, such as the
ZX-calculus~\cite{Coecke2008icalp}, we are directly working on
an explicit representation of states, but we are still able to
prove properties in a scalable way that does not rely on automation,
as one can compose circuits without adding complexity to the proof.



Our proposal combines several components, which are all represented
using dependent and polymorphic types in \coq.
{\em Finite functions}\/ over $n$-tuples of bits can encode a $n$-qubit
  quantum state.
{\em Lenses}\/ are injections between sets of indices, which can be
  used to describe the wiring of quantum circuits 
  in a compositional way. They are related to the lenses used for
  view-update in programming languages and
  databases~\cite{Foster07combinators}.
{\em Currying}\/ of functions representing states, along a lens, provides a direct
  representation of tensor products.
{\em Polymorphism}\/ suffices to correctly apply
  transformations to curried states. We need this polymorphism to behave
  uniformly, which is equivalent to morphisms being natural
  transformations.

Using these components, we were able to provide a full account of
pure quantum circuits in \coq, on top of the \mathcomp{} library,
proving properties from the ground up. We were also
able to prove a number of examples, such as the correctness of the Shor
coding~\cite{Shor1995physrev} (formalized for the first time, albeit
only for an error-free channel at this point),
the Greenberger-Horne-Zeilinger (GHZ) state preparation~\cite{greenberger1989},
a qutrit encoding of the swap gate \cite{Garcia_Escartin_2013},
and the reversed list circuit~\cite{CoqQ-popl2023}.

Our development is available online~\cite{qecc}.

The plan of this article is as follows.
In Section~\ref{sec:prelim}, we provide a short introduction to
quantum states and circuits.
In Section~\ref{sec:lens}, we define lenses.
In Section~\ref{sec:qfocus}, we provide the mathematical definition
of focusing of a circuit through a lens.
In Sections~\ref{sec:gate}~and~\ref{sec:circuit}, we explain the \coq{}
definitions of gates and their composition.
In Section~\ref{sec:proving}, we introduce some lemmas used in
proof idioms that we apply to examples in Section~\ref{sec:examples}.
In Section~\ref{sec:monoid}, we define noncommutative and commutative monoids
of sequential and parallel compositions of gates.
We present related works in Section~\ref{sec:related} before concluding.

\section{Quantum circuits and unitary semantics}
\label{sec:prelim}
In this section, we present basic notions from linear algebra to describe the
unitary model of quantum computation, and how they appear in a quantum circuit diagram.

\subsection{Quantum states}
\label{sec:quantum-states}
Let us first recall that pure classical computation can be seen as a sequence of
boolean functions acting on an array of bits of type $2 ^ n$ for some $n$.
Similarly, pure quantum computation is modeled, in terms of linear algebra, as
a sequence of unitary transformations that act on a quantum state of type $\mathbb C ^ {2 ^ n}$.

A quantum bit (or {\em qubit}) is the most basic unit of data in quantum computation.
We regard it as a variable of type $\mathbb C^2$ and each vector of norm $1$ is considered to be a state of the qubit.
$\mathbb C^2$ has a standard basis $(1, 0), (0, 1)$, which we denote in the context of quantum programming
$\ket{0}, \ket{1}$, indicating that the state of the qubit is $0$ and $1$ respectively.
Regarding $\mathbb C^2$ as the function space $\ord 2 \to \mathbb C$,
where $\ord{n}$ stands for $\{0, \dots, n-1\}$,
we can express the standard basis in the form of functions
\[
\ket 0 := x \mapsto
\begin{cases} 1 & \text{if}\ x = 0 \\ 0 & \text{otherwise} \end{cases}
\quad
\ket 1 := x \mapsto
\begin{cases} 1 & \text{if}\ x = 1 \\ 0 & \text{otherwise} \end{cases}
\]
States other than basis states are linear combinations, which we call {\em superpositions}.
The state of a qubit is mapped to a classical bit by an operation called {\em measurement},
which probabilistically results in values $0$ or $1$.  The measurement of a state in superposition
$a \ket 0 + b \ket 1$ results in $0$ with probability $|a|^2$ and $1$ with probability $|b|^2$.

Those definitions naturally extend to $n$-ary quantum states.
The basis states for $n$ qubits are functions
\[
\ket{i_1i_2\dots i_n} :=
(x : \ord 2^n) \longmapsto
 \begin{cases} 
   1 & \text{if}\ x = (i_1, i_2, \dots, i_n) \\ 0 & \text{otherwise}
 \end{cases}
\]
States other than basis states are again superpositions,
which are linear combinations of norm $1$.  In other words,
a state is represented by a function of type $\mathbb C ^ {2 ^ n}$,
besides the condition on its norm.
We hereafter regard this type as the space of states.
This type can also be identified with the $n$-ary tensor power
 $(\mathbb C^2) ^ {\otimes n}$ of $\mathbb C^2$,
a usual presentation of states in textbooks.

Similarly to the unary case, a measurement of an $n$-ary quantum state
$ \sum_{i\in 2^n} c_i \ket {i_1i_2\dots i_n} $
results in an array of classical bits $i = (i_1, i_2, \dots, i_n)$ with probability
$|c_i|^2$.

\subsection{Unitary transformations}

We adopt the traditional view that pure quantum computation amounts to
applying unitary transformations to a quantum state.
A unitary transformation is a linear function from a vector space to itself
that preserves the inner product of any two vectors, that is,
$\langle U(a) \mid U(b) \rangle$ is equal to $\langle a \mid b \rangle$
for any unitary $U$ and vectors $a$ and $b$,
if we denote the inner product by $\langle a \mid b \rangle$.
Since the norm of $a$ is defined to be $\sqrt{\langle a \mid a \rangle}$,
a unitary also preserves the norm condition of quantum states.

\subsection{Quantum circuits}
\label{sec:quantum-circuits}

\begin{figure}[t]
\scriptsize
\begin{minipage}{0.6\textwidth}
  \scalebox{0.9}{$\shorcode{}$}
  \caption{Shor 9-qubit code}
  \label{fig:shorcode}
\end{minipage}
\quad
\begin{minipage}{0.39\textwidth}
  \scalebox{0.9}{$\bitflipcode$}
  \caption{Bit-flip code}
  \label{fig:bitflipcode}

  \vspace{4ex}

  \scalebox{0.9}{$\signflipcode$}
  \caption{Sign-flip code}
  \label{fig:signflipcode}
\end{minipage}
\end{figure}

In the same way that classical computation can be expressed by an
electronic circuit comprised of boolean gates (AND, OR, etc.),
quantum computation is also conveniently presented as a circuit
with quantum gates that represent primitive unitary transformations.
More generally, a quantum circuit may contain nonunitary operations
such as measurement, but we restrict ourselves to pure quantum circuits
that contain none of them.

A quantum circuit is a concrete representation of quantum computation,
drawn as $n$ parallel wires with quantum gates and other larger subcircuits
being placed over those wires.
A quantum state is input from the left end of a circuit,
transformed by gates and subcircuits on the corresponding wires,
and output from the right end.

As an example, we show the Shor 9-qubit error correction code
(Figure~\ref{fig:shorcode}) and the bit-flip and sign-flip
error correction codes
(Figures~\ref{fig:bitflipcode}~and~\ref{fig:signflipcode}).
Each of the large boxes \framebox{\small${Ch}_n$} denotes an arbitrary
unitary transformations on $n$-qubits, which models a possibly erroneous channel.
The gates placed to the left of \framebox{\small${Ch}_n$} implement
the encoder algorithm of the code, and those to the right the decoder.
The two smaller codes are subcomponents of the Shor code, in the sense
that the encoder and decoder parts of the former appear as subcircuits
in the correposponding parts of the Shor code.
This fact is utilized in the formalization in Section~\ref{sec:examples}.

The primitive operations in a quantum circuit are quantum gates.
One of the simplest gates is the Quantum Not (QNOT) gate
$\footnotesize \Qcircuit @C=1em @R=.7em { & \targ & \qw }$
that operates on one qubit and exchanges the coefficients of
$\ket 0$ and $\ket 1$.
Another simple one is the swap gate
\raisebox{0.5ex}{$\footnotesize \Qcircuit @C=1em @R=.7em { & \qswap & \qw \\ & \qswap \qwx & \qw }$}, which exchanges the states on two wires.
In the Shor code, three kinds of gates appear, namely
the Hadamard
$\footnotesize \Qcircuit @C=1em @R=.7em { & \gate{H} & \qw }$,
the Controlled Not (CNOT)
\raisebox{1.5ex}{$\footnotesize \Qcircuit @C=1em @R=.7em { & \ctrl{1} & \qw \\ & \targ & \qw }$}, and the Toffoli
\raisebox{1.5ex}{\scriptsize $\Qcircuit @C=1em @R=.3em { & \ctrl{2} & \qw \\ & \ctrl{1} & \qw \\ & \targ & \qw }$} gates.

The unitary operations denoted by these gates can be expressed
as matrices with respect to the lexicographically ordered standard basis
(e.g. $\ket {00},\ket {01},\ket {10},\ket {11}$ for two qubits):
\[
\Qcircuit @C=1em @R=.7em { & \targ & \qw }
=
  \begin{bmatrix}
    0 & 1 \\
    1 & 0
  \end{bmatrix}
\qquad
\Qcircuit @C=1em @R=.7em { & \gate{H} & \qw }
=
\frac1{\sqrt 2}
  \begin{bmatrix}
    1 & 1 \\
    1 & -1
  \end{bmatrix}
\]
\[
\raisebox{0.5em}{\Qcircuit @C=1em @R=.7em { & \qswap & \qw \\ & \qswap \qwx & \qw }}
=
  \begin{bmatrix}
      1&0&0&0\\
      0&0&1&0\\
      0&1&0&0\\
      0&0&0&1
  \end{bmatrix}
\quad
\raisebox{1em}{\Qcircuit @C=1em @R=.7em { & \ctrl{1} & \qw \\ & \targ & \qw }}
=
  \begin{bmatrix}
      1&0&0&0\\
      0&1&0&0\\
      0&0&0&1\\
      0&0&1&0
  \end{bmatrix}
\quad
\raisebox{2ex}{\Qcircuit @C=1em @R=.3em { & \ctrl{2} & \qw \\ & \ctrl{1} & \qw \\ & \targ & \qw }}
=
{\scriptsize
  \begin{bmatrix}
      1&0&0&0&0&0&0&0\\
      0&1&0&0&0&0&0&0\\
      0&0&1&0&0&0&0&0\\
      0&0&0&1&0&0&0&0\\
      0&0&0&0&1&0&0&0\\
      0&0&0&0&0&1&0&0\\
      0&0&0&0&0&0&0&1\\
      0&0&0&0&0&0&1&0
  \end{bmatrix}
}
\]


A gate composed in a circuit is represented by a matrix by,
first taking the Kronecker product with identity matrices
corresponding to irrelevant wires, and second sandwiching it
with the matrices that represent the action of a permutation on
the index of tensors to reorder the input and output wires.
For example, to describe the leftmost CNOT gate in the Shor code,
we first \emph{pad} (append) seven wires to CNOT by taking the Kronecker product with
$I_{2^7} = I_{128}$ and apply the permutation $(24)$ to move
$\oplus$ from the second wire to the fourth wire.
The resulting matrix is:
\[
  U_{2^9}((42))
  \begin{bmatrix}
    I_{128}&0&0&0\\
    0&I_{128}&0&0\\
    0&0&0&I_{128}\\
    0&0&I_{128}&0
  \end{bmatrix}
  U_{2^9}((24))
\]
where $U_{2^9}((24))$ denotes the matrix representation of $(24)$
that maps the basis vectors $\ket {i_1\,i_2\,i_3\,i_4\,i_5\,i_6\,i_7\,i_8\,i_9}$ to
$\ket {i_1\,i_4\,i_3\,i_2\,i_5\,i_6\,i_7\,i_8\,i_9}$, and its inverse $U_{2^9}((42))$ is
the same since $(42)=(24)$.

The above method realizes the padding and permutation as linear transformations,
resulting in multiplications of huge matrices.
Taken literally, this method is compositional in that the embedding of
a smaller circuit into a larger one can be iterated,
but impractical because of the exponential growth of the dimension of
the matrices.
A way to avoid this problem is to stick to a symbolic representation
based on sums of matrix units, that can ignore zero components, but it
is less compositional, in that the representation of the gate is
modified to fit an application site, leading to different representations
and reasoning at different sites.
We aim at solving this problem by separating the wiring part,
which is a combinatorics that does not essentially touch quantum states,
from the actions of a quantum gate, which is an intrinsic property of
the gate itself.


\section{Lenses}
\label{sec:lens}
The first element of our approach is to provide a data structure,
which we call a {\em lens}, that describes the composition of a
subcircuit into a circuit.
It forms the basis for a combinatorics of composition.

\subsection{Lenses in programming}
\label{sec:lenses-in-programming}
\newcommand\Get{\textbf{get}}
\newcommand\Put{\textbf{put}}
The concept of lens~\cite{Foster07combinators} was introduced in the
programming language
community as a way to solve the {\em view-update}
problem~\cite{bancilhon1981transdb}, which itself comes from the
database community. Lenses are often described as a pair of
functions $\Get : S \to V$ and $\Put : V \times S \to S$,
where $S$ denotes the whole state and $V$ a partial view on it,
and which satisfy the following laws.
\[ \begin{array}{rcl}
\textsc{GetPut} &:& \Put(\Get(s),s) = s, \\
\textsc{PutGet} &:& \Get(\Put(v,s)) = v.
\end{array} \]

A more versatile approach adds the concept of complementary
view~\cite{bancilhon1981transdb,Barbosa2010icfp},
which adds another type $C$ and a function
$\lensC{\Get} : S \to C$,
changing the type of $\Put$ to $V \times C \to S$,
so that the laws become:
\[ \begin{array}{rcl}
\textsc{GetPut} &:& \Put(\Get(s),\lensC{\Get}(s)) = s, \\
\textsc{PutGet} &:& \Get(\Put(v,c)) = v, \\
\textsc{PutGetC} &:& \lensC{\Get}(\Put(v,c)) = c. \\
\end{array} \]

Our representation of lenses is an instance of the second approach.
This choice is necessary from a quantum computational viewpoint.
The first approach considers the projective copy of data from $S$ when using
a \Get, and the overwriting of data (and hence partial deletion) in $S$ when
using a \Put. Both these operations are irreversible,
and hence restricted in quantum computing.
The second approach can be used to define quantum lens without such problems.

\subsection{Combinatory lenses}
\begin{figure}
\[ \begin{array}{rcl}
\lens{n}m &\cong& \ord{m} \rightarrowtail \ord{n} \\
\coqin{extract}_{T,n,m} &:& \lens{n}m \to \tuple{T}n \to \tuple{T}m \\
\coqin{merge}_{T,n,m} &:& \lens{n}m \to \tuple{T}m \to \tuple{T}{n-m}
\to \tuple{T}n \\
\coqin{lensC}_{n,m} &:& \lens{n}m \to \lens{n}{n - m} \\
\_ \lcomp_{n,m,p} \_ &:& \lens{n}m \to \lens{m}p \to \lens{n}p \\
\coqin{lens_basis}_{n,m} &:& \lens{n}m \to \lens{n}m \\
\coqin{lens_perm}_{n,m} &:& \lens{n}m \to \lens{m}m \\
\end{array} \]
\caption{Lenses, actions, and operations}
\label{fig:lens}
\end{figure}

We want to map the $m$ wires of a subcircuit to the $n$ wires of the
external one. This amounts to defining an injection from $\ord{m}$ to
$\ord{n}$, on top of which we can define the actions and operations of
Figure~\ref{fig:lens}.

Throughout this article,
we use mathematical notations to make our \coq{}
code easier to read. For instance $\ord{n}$ in the above
definition of lens denotes the ordinal type
\coqin{'I_n} of \mathcomp, and $\tuple{T}m$ denotes the type of
tuples of arity $m$ with elements of type $T$
(i.e. the type \coqin{m.-tuple T}).
We write type parameters as indices, and allow for omitting them.
When showing lemmas, we also use the keyword \coqin{Variables}, which
is part of the section mechanism of \coq, in a relatively loose
manner. It denotes variables that are common to the following
definitions. The end of its scope should be clear from the context.

We call \emph{focusing} the operation using a lens to update a system
according to changes in a subsystem.
The \coqin{extract}, \coqin{merge} and \coqin{lensC} operations in
Figure~\ref{fig:lens} are basic and required to define focusing.
The \Get{} operation of a lens $\ell$ is $\coqin{extract}\ \ell$,
which is the projection of $\tuple{T}n$ onto $\tuple{T}m$ along
$\ell$, {\it i.e.} it returns a tuple containing the elements whose
indices appear in $\mathrm{Im}(\ell)$, in the same order as $\ell$.
Each lens $\ell$ has its complementary lens $\coqin{lensC}\ \ell$, which is the
unique monotone bijection from $\ord{n-m}$ to
$\ord{n} \smallsetminus \mathrm{Im}(\ell)$.
We will write $\lensC\ell$ for $\coqin{lensC}~\ell$.
Their composition $\coqin{extract}\ \lensC\ell$ returns the complementary view.
The corresponding \Put{} operation is $\coqin{merge}\ \ell\ v\ c$,
which internally uses $\lensC\ell$; namely, for each position $i$ of the
output tuple, if this position appears in $\mathrm{Im}(\ell)$, it returns the
value at index $\ell^{-1}(i)$ in $v$; otherwise, this position must appear
in $\mathrm{Im}(\lensC\ell)$, and it returns the value at
index ${\lensC\ell}^{-1}(i)$ in $c$.

As an example, let us consider the following lens:
\[ \ell : \lens{4}2 := \{0\mapsto 2, 1 \mapsto 0\}. \]
Then we have:
\[ \begin{array}{rcl}
\coqin{extract}~\ell~[a,b,c,d] &=& [c,a] \\
\coqin{merge}~\ell~[a,b]~[c,d] &=& [b,c,a,d] \\
\lensC\ell &=& \{0\mapsto 1, 1 \mapsto 3\}
\end{array} \]

In the following, the lens $\ell$ will be available from the context, so that
we omit it in \coqin{extract} and \coqin{merge}, and
$\extractC$ denotes $\coqin{extract}\ \lensC\ell$.
The \textsc{GetPut}, \textsc{PutGet} and \textsc{PutGetC} laws become:
\begin{minted}{ssr}
Variables 7$(n\ m : \Nat)\ (\ell : \lens{n}m)\ (T : \texttt{Type})\ (s : \tuple{T}n)\ (v : \tuple{T}m)\ (c : \tuple{T}{n-m})$7.
Lemma merge_extract  : merge (extract 7$s$7) (7$\extractC\ s$7) = 7$s$7.
Lemma extract_merge  : extract (merge 7$v\ c$7) = 7$v$7.
Lemma extractC_merge : 7$\extractC$7 (merge 7$v\ c$7) = 7$c$7.
\end{minted}

Let us describe the remaining operations of Figure~\ref{fig:lens}.
It is often useful to compose lenses,
i.e. given two lenses $\ell_1 : \lens{n}m$ and $\ell_2 : \lens{m}p$,
their composition $\ell_1 \lcomp \ell_2$ has type $\lens{n}p$.
One can use composition to factorize a lens into its basis (the
monotone part) and permutation part.
\begin{center}
  \begin{tikzcd}[row sep=1em]
    \ord{m} \arrow[rd, tail, two heads, "\textrm{perm.}"'] \arrow[rr, tail, "\ell"] &  & \ord{n} \\
    & \ord{m} \arrow[ru, tail, "\textrm{basis, monotone}"'] &
  \end{tikzcd}
\end{center}
Namely, we have the following laws:
\begin{minted}{ssr}
Lemma lens_basis_perm : 7$(\texttt{lens\_basis}\ \ell) \lcomp (\texttt{lens\_perm}\ \ell) = \ell$7.
Lemma mem_lens_basis  : 7$\mathrm{Im}(\texttt{lens\_basis}\ \ell) =_i \mathrm{Im}(\ell)$7.
\end{minted}
where $\mathrm{Im}(\ell_1) =_i \mathrm{Im}(\ell_2)$ means that
$\ell_1$ and $\ell_2$ have same images as sets.

Reusing our example above, the basis and permutation of $\ell$ are:
\[ \begin{array}{rcl}
\coqin{lens_basis}~\ell &=& \{0\mapsto 0, 1 \mapsto 2\} \\
\coqin{lens_perm}~\ell &=& \{0\mapsto 1, 1 \mapsto 0\}
\end{array} \]

\subsection{Classical focusing}
We show the classical case of focusing (\coqin{focus1}) as an example
(Figure~\ref{fig:classical_lens}, page~\pageref{fig:classical_lens}).
In this case, data is represented by direct products, whose
elements are tuples, readily manipulated by \coqin{extract} and
\coqin{merge}. A change on the subsystem of type
$\tuple{T}m$ is thus propagated to the global state of type $\tuple{T}n$.
\begin{minted}{ssr}
Definition focus17$_{T,n,m}\ (\ell : \lens{n}m)\ (f : \tuple{T}m \to \tuple{T}m)$7 : 7$\tuple{T}n \to \tuple{T}n$7 :=
  7$s\ \ \mapsto \texttt{merge}\ (f\ (\texttt{extract}\ s))\ (\extractC\ s)$7.
Variables (7$T$7 : Type) 7$(n\ m\ p : \Nat)\ (\ell : \lens{n}p)\ (\ell_1 : \lens{n}m)\ (\ell_2 : \lens{m}p)$7
          7$(f : \tuple{T}p \to \tuple{T}p)$7.
Lemma focus1_in : 7$\texttt{extract}\ \ell \circ (\texttt{focus1}\ \ell\ f) = f \circ \texttt{extract}\ \ell$7.
Lemma focus1A : 7$\texttt{focus1}\ \ell_1\ (\texttt{focus1}\ \ell_2\ f) = \texttt{focus1}\ (\ell_1\lcomp\ell_2)\ f$7.
\end{minted}
The function \coqin{focus1} first decomposes the state $s$ into the view
$\texttt{extract}\ s$ and
its complement, then applies $f$ to the view, and combines it with the
complement to return the updated state.
The lemma \coqin{focus1_in} states that extracting after applying a
focused function (using the same lens), is the same as applying this
function after extracting.
And lemma \coqin{focus1A} says that \coqin{focus1} is a morphism for
lens composition.

\coqin{focus1} cannot be directly applied to quantum state
transformations,
where the state is not represented by direct products but by tensor
products.
We will see in the next sections
that its quantum version can be defined through currying
and uncurrying of quantum states, which can both be in turn defined
using the three previous operations.
Note that the use of a complementary view is important too, as
no-cloning theorems mean that we cannot duplicate information.

\subsection{Implementation}
The lens data structure and the functions and lemmas explained in this
whole section can be found in the \coqin{lens} module of our \coqin{qecc}
library~\cite[\coqin{lens.v}]{qecc}.

In our implementation, we choose to represent canonically a lens as
an $m$-tuple of indices in $\ord{n}$, without repetition, as this allows
direct computation.
\begin{minted}{ssr}
Record 7$\lens{n}m := \texttt{mkLens} \left\{ \ell : \tuple{\ord{n}}m \mid \texttt{uniq}\ \ell\right\}$7.
\end{minted}
The above definition is a dependent record.
It defines a subtype of the tuple type $\tuple{\ord{n}}m$, such that
values of type $\lens{n}m$ can be automatically coerced to
$\tuple{\ord{n}}m$. Given a tuple $t : \tuple{\ord{n}}m$ and a proof
$H : \texttt{uniq}\ t$, one can build a new lens by writing
$\tt mkLens\ H$, where $t$ is given implicitly through $H$.
This follows the same pattern as ordinals $\ord{n}$ and tuples
$\tuple{T}m$, from the \mathcomp{} library:
\begin{minted}{ssr}
Record 7$\ord{n} := \texttt{Ordinal} \left\{m : \Nat \mid m < n\right\}$7.
Record 7$\tuple{T}n := \texttt{Tuple} \left\{s : \texttt{seq}\ T \mid \texttt{size}\ s\ \texttt{==}\ n\right\}$7.
\end{minted}

The definition of \coqin{extract} is easy, as it just amounts to mapping
over the tuple representation of lenses.
Hereafter $t[i]$ denotes the $i^{\rm th}$ element of the tuple $t$, aka
$\coqin{tnth}\ t\ i$.
\begin{minted}{ssr}
Definition extract7$_{T,n,m}\ (\ell : \lens{n}m)\ (t : \tuple{T}n) :\tuple{T}m$7 := map_tuple 7$(i \mapsto t[i])\ \ell$7.
\end{minted}

Due to the dependent typing of lenses, the definition of
$\coqin{lensC}$ is more complex.
In our library, it is done in two steps. First obtain the ordered list
of the missing indices ($\texttt{enum}\ \ord{n}$ is the ordered list
of all values of type $\ord{n}$, which we filter with the condition):
\begin{minted}{ssr}
Definition seq_lensC7$_{n,m}\ (\ell : \lens{n}m)\ $7:= [seq 7$i \leftarrow {\tt enum}\ \ord{n} \mid i \notin \ell$7].
\end{minted}
Then, we need two proofs, first about the length of the resulting
list, to turn it into a tuple, then about the absence of repetition in
it, to turn it into a lens.
\begin{minted}{ssr}
Variables 7$(n\ m : \Nat)\ (\ell : \lens{n}m)$7.
Lemma size_lensC : size (seq_lensC 7$\ell$7) == 7$n - m$7.
Lemma uniq_lensC : uniq (Tuple (size_lensC 7$\ell$7)).
Definition lensC7$_{n,m}\ (\ell : \lens{n}m) : \lens{n}{n-m}$7 := mkLens (uniq_lensC 7$\ell$7).
\end{minted}
We will follow the same pattern of lemma followed by a definition
relying on it for other definitions of dependently typed values.

The implementation of \coqin{merge} introduces a different problem.
A weakly typed implementation uses indexing in the list structure of lenses.
\begin{minted}{ssr}
Definition merge_nth7$_{I,n,m}\ (d_I : I)\ (\ell : \lens{n}m)\ (v : \tuple{I}m) (w : \tuple{I}{n-m})$7 :=
    [tuple nth (nth 7$d_I\ w$7 (index 7$i\ \lensC\ell$7)) 7$v$7 (index 7$i\ \ell$7) | 7$i < n$7].
\end{minted}
This definition relies on the respective exceptional behaviors of
\coqin{index} and \coqin{nth}. Namely, $\coqin{index}~i~\ell$ will
return $m$ if $i$ is not in $\ell$, and $\coqin{nth}~d~v~i$ will
return $d$ if $i$ is greater or equal to the length of $v$. As a
result, this definition returns an element of $v$ if $i$ is in $\ell$,
and element of $w$ if $i$ is in $\lensC\ell$, and $d_I$ otherwise. This
last case cannot happen, but this requires a separate proof.

A more elegant definition relies on the notion of \textit{lens index},
which is a new dependently typed value, realizing a partial inverse of
the lens injection.
\begin{minted}{ssr}
Variables 7$(n\ m : \Nat)\ (\ell : \lens{n}m)\ (i : \ord{n})$7.
Lemma index_tuple7$_I :\forall t : \tuple{I}n,\ i \in t \to \texttt{index}\ i\ t < n$7.
Definition lens_index 7$(H : i \in \ell) : \ord{m} := \texttt{Ordinal}\ (\texttt{index\_tuple}\ H)$7.
Lemma tnth_lens_index : 7$\forall (H : i \in \ell),\ \tnth\ell{\lindex H} = i$7.
Lemma mem_lensC : 7$(i \in \lensC\ell) = (i \notin \ell)$7.
Lemma mem_lensFC : 7$i \in \ell = {\tt false} \to i \in \lensC\ell$7.
\end{minted}
Given a proof $H$ that $i\in\ell$, the function
$\coqin{lens_index}~H$ returns the unique index $j : \ord{m}$ such
that $\ell[j] = i$.
It can be used to define \coqin{merge} through a dependent case
analysis on the boolean $i\in\ell$, using the lemma \coqin{mem_lensFC}
to turn the false case into membership in $\lensC\ell$.
\begin{minted}{ssr}
Definition merge7$_{I,n,m}\ (\ell : \lens{n}m)\ (v : \tuple{I}m)\ (w : \tuple{I}{n-m})$7 :=
    [tuple match sumbool_of_bool (7$i \in \ell$7) with
           | left  7$e \Rightarrow v[\lindex{e}]$7
           | right 7$e \Rightarrow w[\lindex{(\texttt{mem\_lensFC}\ e)}]$7 end
    | 7$i < n$7].
\end{minted}
Both definitions have advantages and inconveniences. \coqin{merge_nth}
is computable, and allows one to get rid of dependent types.
On the other hand, \coqin{merge} is more abstract, and encourages
cleaner proofs, using the following two lemmas.
Given a default value,
we can still fall back to \coqin{merge_nth} when reduction is needed.
\begin{minted}{ssr}
Variables (7$I$7: Type) 7$(v : \tuple{I}m)\ (w : \tuple{I}{n-m})$7.
Lemma tnth_merge : 7$\forall(H : i \in \ell),\ \tnth{(\texttt{merge}\ v\ w)}i = \tnth{v}{\lindex H}$7.
Lemma tnth_mergeC : 7$\forall(H : i \in \lensC\ell),\ \tnth{(\texttt{merge}\ v\ w)}i = \tnth{w}{\lindex H}$7.
Lemma mergeE : 7$\forall d_I\ v\ w,\ \texttt{merge}\ \ell\ v\ w = \texttt{merge\_nth}\ d_I\ \ell\ v\ w$7.
\end{minted}

The notion of lens index is versatile, and it is also a good fit with
lens composition.
\begin{minted}{ssr}
Variables 7$(n\ m\ p : \Nat)\ (\ell_1 : \lens{n}m)\ (\ell_2 : \lens{m}p)$7.
Lemma lens_comp_uniq : uniq (extract 7$\ell_2\ \ell_1$7).
Definition 7$\ell_1 \lcomp_{n,m,p} \ell_2 : \lens{n}p$7 := mkLens (lens_comp_uniq 7$\ell_1\ \ell_2$7).
Lemma mem_lens_comp : 7$\forall i\,(H : i \in \ell_1),\ (i \in \ell_1 \lcomp \ell_2) = (\texttt{lens\_index}\ H \in \ell_2)$7.
\end{minted}
Lens composition defining a new lens, it is again a dependently typed
value. The tuple part amounts just to extracting the $\ell_2$ view
from $\ell_1$. And an essential property of lens composition is that
membership in $\ell_1 \lcomp \ell_2$ can be expressed as membership of
a lens index on $\ell_1$ in $\ell_2$.

\subsection{Application to compact data structures}
There is an interesting connection between our definition of
\coqin{lensC} and the  $\coqin{select}_0$ function commonly used for
compact data structures~\cite{navarro2016}.

Namely, assuming a bitstring $s$, the
$\coqin{rank}_0(s,i)$ function counts the number of elements at value
0 in the first $i$ bits of $s$.
\[ \coqin{rank}_0(s,i) =
   \left| \left\{ j < i \mid s[j] = 0 \right\} \right| \]
Conversely the $\coqin{select}_0(s,i)$ function returns the position
of the $i^{\rm th}$ bit of value 0 in $s$, indices starting from 1 (if
$i=0$, it conventionally returns 0,
and if there are less than $i$ bits at value 0 in $s$, it returns the
length of $s$ plus 1). If $\coqin{rank}_0(s) \geq i$, it can also be defined as
\[ \coqin{select}_0(s,i) = \min\{j \mid \coqin{rank}_0(s,j) \geq i\} \]

Now, if for $s$ we take the characteristic
function $\coqin{bits}(\ell)$ of the lens $\ell$ of type $\lens{n}m$,
and $\coqin{bits}(\ell)$ contains no less than $i > 0$ bits of value 0,
then the following equalities stand:
\[ \begin{array}{rcl}
\coqin{bits}(\ell) &=& [{\tt seq} ~(i \in \ell) \mid i \leftarrow {\tt
    enum}\ord{n}] \\
\coqin{select}_0(\coqin{bits}(\ell),i) &=& \lensC\ell[i-1] + 1
\end{array} \]
The change in indices comes from the 1-based indexing in $\coqin{select}_0$.

\section{Quantum lenses}
\label{sec:qfocus}

\begin{figure}[t]
  \small
\begin{minipage}{0.4\textwidth}
  \begin{tikzcd}
    S = T^n \arrow[rrr, "\coqin{extract}\ \ell"] \arrow[d, xshift=1.5ex, "\coqin{focus1}\ \ell"] &
    \arrow[phantom]
    \arrow[d, no head, to path={
      ([xshift=2em]\tikztostart.center)
      .. controls +(0.7,0) and +(0.7,0) ..
      ([xshift=2em]\tikztotarget.center)
    }]
    & \arrow[phantom] & T^m = V \arrow[d, xshift=-2.5ex, "f"]\\
    \phantom{S =} T^n & \arrow[phantom]
    & \arrow[phantom] & \arrow[lll, "\coqin{merge}\ \ell"] T^m \phantom{= V}
  \end{tikzcd}
  \caption{Classical focusing}
  \label{fig:classical_lens}
\end{minipage}
\quad
\begin{minipage}{0.55\textwidth}
  \begin{tikzcd}
    S = T^{2^n} \arrow[rrr, "\coqin{curry}\ \ell"] \arrow[d, xshift=1ex, "\coqin{focus}\ \ell"] &
    &  & (T^{2^{n-m}})^{2^m} = V \arrow[d, xshift=-3ex, "G_{T^{2^{n-m}}}"]
    & T^{2^m} \arrow[d, xshift=-1ex, "G_T"]
   \\
    \phantom{S =} T^{2^n} & 
    &  & \arrow[lll, "\coqin{uncurry}\ \ell"] (T^{2^{n-m}})^{2^m} \phantom{= V}
    & T^{2^m}
  \end{tikzcd}
  \caption{Quantum focusing}
  \label{fig:quantum_lens}
\end{minipage}
\end{figure}

In this section, we give a mathematical account for
actions of lenses (i.e., the \Get{} and \Put{} operators in
Section~\ref{sec:lenses-in-programming})
on quantum states and operators, and a quantum variant of focusing
that facilitates the composition of quantum gates as sketched in
Section~\ref{sec:quantum-circuits}.
The classical \Get{} and \Put{} introduced in the previous section
(the operators \coqin{extract} and \coqin{merge})
play an important role in the definition of quantum ones.
The definitions and lemmas in this section can be found in the
\coqin{dpower} module~\cite[\coqin{dpower.v}]{qecc}.

\subsection{Curry and uncurry}

We first define the quantum \Get{} and \Put{} operators.
Recall that the type of $n$-qubit quantum states was defined as
a double power $\mathbb C ^ {2 ^ n}$ in Section~\ref{sec:quantum-states}.
The quantum \Get{} and \Put{} should allow one to select a few qubits
out of a given state and apply a quantum operator on them.
A crucial difference in the quantum case is that \Get{}
must not discard the irrelevant qubits,
unlike the classical one that was defined as a projection.

Such a quantum \Get{} and the corresponding \Put{}
can be defined in a form of currying and uncurrying
that constitute an isomorphism
$
  T^{2^n}
  \left(\cong T^{2^{n-m} 2^m}\right)
  \cong \left( T^{2^{n-m}} \right)^{2^m}
$:
\begin{minted}{ssr}
Definition curry7$_{T,n,m}\ \ \  : \lens{n}m \to T^{2^n} \to (T^{2^{n-m}})^{2^m}$7.
Definition uncurry7$_{T,n,m} : \lens{n}m \to (T^{2^{n-m}})^{2^m} \to T^{2^n}$7.
\end{minted}
The type parameter $T$ is generalizing $\mathbb C$ in the definition
of qubits. It is intended to vary over $\mathbb C$-modules
(in the sense of modules over a commutative ring),
including $\mathbb C$ itself as a one-dimensional free $\mathbb C$-module.
For a lens $\ell$ of type $\lens n m$,
the result of applying $\coqin{curry}\ \ell$ to an input state $\sigma \in T^{2^n}$
is a function that takes two indexing tuples $v \in 2^m$ and $w \in 2^{n-m}$
and returns the evaluation of $\sigma$ at $\coqin{merge}\ \ell\ v\ w$,
the combined index of $v$ and $w$ along $\ell$.
Its inverse $\coqin{uncurry}\ \ell$ is defined similarly
for $\sigma \in (T^{2^{n-m}})^{2^m}$ and $v \in 2^n$,
by extracting two indices $\coqin{extract}\ \ell\ v$ and $\extractC\ \ell\ v$
from $v$ along $\ell$.
\begin{align*}
&\coqin{curry}\ \ell\ \sigma\ v\ w := \sigma\ (\coqin{merge}\ \ell\ v\ w)\\
&\coqin{uncurry}\ \ell\ \sigma\ v := \sigma\ (\coqin{extract}\ \ell\ v)\ (\extractC\ \ell\ v)
\end{align*}
The fact that \coqin{curry} and \coqin{uncurry} indeed form an isomorphism is observed as
cancellation lemmas, which are derived from the laws
\coqin{merge_extract}, \coqin{extract_merge}, \coqin{extractC_merge}:
\begin{minted}{ssr}
Lemma curryK   : uncurry7$\ \ell$7 7$\circ$7 curry7$\ \ell$7 7$= \textrm{id}_{T^{2^n}}$7.
Lemma uncurryK : curry7$\ \ell$7 7$\circ$7 uncurry7$\ \ell$7 7$= \textrm{id}_{(T^{2^{n-m}})^{2^m}}$7.
\end{minted}

\subsection{Quantum focusing}

We proceed to defining composition of quantum gates in a circuit by means of
\coqin{curry} and \coqin{uncurry}.
An $m$-qubit quantum gate $G$ is a linear transformation on $\mathbb C ^{2^m}$,
and it can be represented by a square matrix of type $\mathcal M_{2^m}(\mathbb C)$.
The action of such a matrix $M$ on a $2^m$-dimensional vector is computed
by scalar multiplications and additions,
hence it can be extended for the direct power $T^{2^m}$ of any $\mathbb C$-module $T$:
\begin{align}
  &Mv = \sum_{1 \leq j \leq 2^m} (M_{(1,j)}v_j , \dots, M_{(2^m,j)}v_j)^t \label{eq:M-action}\\
  &\text{(where $M = (M_{(i,j)})$, $v = (v_j)$, and each $v_j \in T$)} \notag
\end{align}

This freedom in the choice of $T$ can be captured by endowing $G$ the following
polymorphic type of linear transformations.
\begin{equation}
  \label{eq:G-polytype}
  G : \forall T : \textrm{$\mathbb C$-module},
  T^{2^m}
  \stackrel{\mathrm{linear}}\longrightarrow
  T^{2^m}
\end{equation}

Along with the \coqin{curry}-\coqin{uncurry} isomorphism above,
a gate $G$ can be applied to
a larger number of qubits, to become composable in a circuit.
This realizes quantum focusing (Figure~\ref{fig:quantum_lens}).
\[
  \coqin{focus}\ \ell\ G :=
  {\Lambda\,T.}
  (
  \coqin{uncurry}\ \ell \circ
  G_{T^{2^{n-m}}} \circ
  \coqin{curry}\ \ell
  )
\]

So far, the type of $G$ (Equation~\ref{eq:G-polytype}) has told that each instance
$G_T$ is linear and
can be represented by a matrix, but not yet that they are the same matrix
for any $T$.
We impose the uniqueness of the matrix as an additional property as follows.
\[
  \exists M : \mathcal M_{2^m}(\mathbb C),\ \forall T : \textrm{$\mathbb C$-module},\ %
  \forall v : T^{2^m},\  G_T (v) = M v.
\]
The multiplication $Mv$ is defined as in Equation~\ref{eq:M-action}.
This existence of a unique matrix representation implies the uniformity of
the actions of $G$, which amounts to commutativity of the following diagram.
\[\begin{tikzcd}[row sep=1em]
    T & {T^{2^m}} && {T^{2^m}} \\
    \\
    {T'} & {T'^{2^m}} && {T'^{2^m}}
    \arrow["\forall\varphi"', from=1-1, to=3-1]
    \arrow["{\varphi^{2^m}}"', from=1-2, to=3-2]
    \arrow["{\varphi^{2^m}}", from=1-4, to=3-4]
    \arrow["{G_T}", from=1-2, to=1-4]
    \arrow["{G_{T'}}"', from=3-2, to=3-4]
  \end{tikzcd}\]
Importing category-theoretic terminology, we call this property \emph{naturality}:
$G$ is natural with respect to the functor $(-)^{2^m}$.

We proved conversely that this naturality implies the uniqueness of the matrix.
We shall incorporate naturality, instead of an explicit matrix,
in our formal definition of quantum gates in the next section.

\section{Defining quantum gates}
\label{sec:gate}

Using \mathcomp{}, we can easily present the concepts described in the
previous sections.
From here on, we fix $K$ to be a commutative ring.
\mathcomp{} distinguishes a commutative ring $K$ from the
one-dimensional free module over itself (denoted $\coqin{K^o}$ in
\coq{} code), but we will
not make this distinction in this article.

We first define quantum states as the double power $T^{2^n}$
discussed in Section~\ref{sec:prelim}.
It is encoded as a function type $\dpower{T}n$ from $n$-tuples
of some finite type $I$ to a type $T$.
For qubits, we shall have $I = \ord{2} = \{0,1\}$, but we can also naturally
represent qutrits (quantum information units with three states) by
choosing $I = \ord{3}$.
\begin{minted}{ssr}
Variables 7$(I : \textrm{finite type})\ (K : \textrm{commutative ring})\ (T : K\textrm{-module})$7.
Definition 7$\dpower{T}n := I^n \xrightarrow{\rm finite} T$7.
Definition dpmap7$_{m,T_1,T_2}\ (\varphi : T_1 \to T_2)\ (s : \dpower{T_1}m) : \dpower{T_2}m$7 := 7$\varphi \circ s.$7
\end{minted}
This construction, $\dpower{(-)}n$,
can be regarded as a functor with its action on functions
provided by \coqin{dpmap}, that is,
any function $\varphi : T_1 \to T_2$ can be extended to
${\tt dpmap}\ \varphi : \dpower{T_1}n \to \dpower{T_2}n$, which are
drawn as the vertical arrows in the naturality square in the previous section.

We next define quantum gates as  natural transformations
(or {\em morphisms}).
\begin{minted}{ssr}
Definition morlin7$_{m,n} := \forall(T : \text{$K$-module}),\ \dpower{T}m \xrightarrow{\rm linear} \dpower{T}n$7.
Definition naturality7$_{m,n}\ (G : \morlin{m}{n})$7 :=
    7$\forall(T_1\ T_2 : \text{$K$-module}),\forall(\varphi : T_1 \xrightarrow{\rm linear} T_2),$7
           7$(\texttt{dpmap}\ \varphi)\circ (G\ T_1) = (G\ T_2)\circ (\texttt{dpmap}\ \varphi)$7.
Record mor7$_{m,n} := \left\{ G : \morlin{m}{n} \mid \texttt{naturality } G\right\}$7.
Notation endo7$_n := (\mor{n}{n})$7.
Definition unitary_mor7$_{m,n}\ (G : \mor{m}{n})$7 := 7$\forall s, t, \langle G_K\ s \mid G_K\ t \rangle = \langle s \mid t\rangle$7.
\end{minted}
A crucial fact we rely on here is that,
for any $K$-module $T$,
\mathcomp{} defines the $K$-module of the finite functions
valued into it, so that $\dpower{T}n$ is a $K$-module.
This allows us to define the type \coqin{morlin} of polymorphic linear functions
between $\dpower{T}m$ and $\dpower{T}n$, and further combine it with naturality
into the types $\coqin{mor}_{m, n}$ of morphisms from
$\dpower{(-)}m$ to $\dpower{(-)}n$ and $\coqin{endo}_n$ of endo-morphisms.

We leave unitarity as an independent property, called
\coqin{unitary_mor}, since it
makes sense to have non-unitary morphisms in some situations.

While most definitions and lemmas in this section are in the
\coqin{dpower} module~\cite[\coqin{dpower.v}]{qecc}, definitions and
lemmas related to unitarity are in the \coqin{unitary}
module~\cite[\coqin{unitary.v}]{qecc}.

Concrete quantum states can be expressed directly as functions in
$\dpower{K}{n}$, or as a linear combination of
computational basis vectors $\ket{v}$, where $v : I^n$ is the
index of the only 1 in the vector.
\begin{minted}{ssr}
Definition 7$\ket{v} : \dpower{K}n$7 := 7$\displaystyle(v' : \tuple{I}n) \finmapsto \left\{\begin{array}{l} 1 \quad \mathrm{if}\ v = v' \\ 0 \quad \mathrm{otherwise} \end{array}\right.$7
\end{minted}
For a concrete tuple, we also write $\ket{i_1, \dots, i_n}$ for
$\ket{[{\tt tuple}~ i_1; \dots; i_n]}$.
This representation of states allows us to go back and forth between
computational basis states and indices, and is amenable to proofs.

Using this basis, one can also define a morphism from its matrix
representation (expressed as a nested double power, in column-major
order, i.e. as a row vector of column vectors).
The reason for this choice is that one often characterizes a
transformation by its image on computational basis states. Each of the
resulting states are column vectors, indexed by the state
they come from.
Here we define the QNOT and CNOT gates as mappings from computational basis indices to
column vectors, using the addition modulo $2$ (written $\oplus$).
The swap gate is similar.
Alternatively, the expression $\coqin{ket_bra k b}$ stands for the product of a column vector and a row vector,
resulting in an $m \times n$ matrix (written $\ket k \bra b$ in the Dirac notation).
We use it to define the Hadamard gate as
a sum of matrix units. Both matrices are then fed to
\coqin{dpmor} to obtain morphisms.
All of these gates work on qubits, and are defined after instantiating
the double power construction~\cite[\coqin{qexamples_common.v}]{qecc}.
\newcommand\ketbra[2]{\texttt{ket\_bra}\,\ket{#1}\,\ket{#2}}
\begin{minted}{ssr}
Definition dpmor7$_{m,n} : \dpower{(\dpower{K}n)}m \to \mor{m}{n}$7.
Definition ket_bra7$_{m,n} \ (k : \dpower{K}m)\ (b : \dpower{K}n) : \dpower{(\dpower{K}n)}m$7 := 7$v \finmapsto (k\ v) \cdot b$7.
Definition qnot : endo7$_1$7 := dpmor 7$(v : \ord{2} \finmapsto \ket{1 \oplus v})$7.
Definition cnot : endo7$_2$7 := dpmor 7$(v : \ord{2}^2 \finmapsto \ket{\tnth{v}0, \tnth{v}0 \oplus \tnth{v}1})$7.
Definition swap : endo7$_2$7 := dpmor 7$(v : \ord{2}^2 \finmapsto \ket{\tnth{v}1, \tnth{v}0})$7.
Definition hadamard : endo7$_1$7 :=
 dpmor7\(\displaystyle\left(\frac{1}{\sqrt 2}\left(\ketbra{0}{0}+\ketbra{0}{1}+\ketbra{1}{0}-\ketbra{1}{1}\right)\!\right)\)7.
\end{minted}

As explained in Section~\ref{sec:qfocus}, naturality for a morphism is
equivalent to the existence of a uniform matrix representation.
\begin{minted}{ssr}
Lemma naturalityP : naturality 7$G\ \longleftrightarrow\ \exists M,\ \forall T,s,\ G_T\ s = (\texttt{dpmor}\ M)_T\ s$7.
\end{minted}
On the right hand side of the equivalence we use the extensional
equality of morphisms, which quantifies on $T$ and $s$. By default, it
is not equivalent to \coq{}'s propositional equality; however the two
coincide if we assume functional extensionality and proof irrelevance,
two relatively standard axioms inside \coq.
\begin{minted}[fontsize=\normalsize]{ssr}
Lemma morP : 7$\forall(F, G : \mor{m}{n}),$7 7$(\forall T, s,\ F_T\ s = G_T\ s) \longleftrightarrow F = G$7.
\end{minted}
While our development distinguishes between the two equalities, in
this article we will not insist on the distinction, and just abusively
write $F = G$ for extensional equality too.
Only in Section~\ref{sec:monoid} will we use those axioms to
prove and use the above lemma.

One can also use naturality to prove the following lemma, which states
that two morphisms are extensionally equal if they instances at $K$
coincide on basis states.
\begin{minted}{ssr}
Lemma eq_mor_basis :
  7$\forall(F, G : \mor{m}{n}), (\forall v,\ F_K\ \ket{v} = G_K\ \ket{v}) \to (\forall T, s,\ F_T\ s = G_T\ s)$7.
\end{minted}
This last lemma does not require any axiom.

\section{Building circuits}
\label{sec:circuit}

The currying defined in Section~\ref{sec:qfocus} allows us to compose
circuits without referring to a global set of qubits.
This is obtained through two operations: (sequential) composition of
morphisms, which just extends function composition, and focusing
through a lens, which allows us to connect the wires of a gate into a
larger circuit.
Again the definitions and lemmas in this section are from the
\coqin{dpower} module~\cite[\coqin{dpower.v}]{qecc}.
\begin{minted}{ssr}
Definition  7${\vcomp_{n,m,p}}\quad  : \mor{m}{p} \to \texttt{mor}_{n,m} \to \texttt{mor}_{n,p}$7.
Definition focus7$_{n,m} : \lens{n}m \to \texttt{endo}_m \to \texttt{endo}_n$7.
\end{minted}
To define \coqin{focus}, we combine currying and polymorphism into
\coqin{focuslin} as we did in Section~\ref{sec:qfocus}, and add a proof of naturality.
\begin{minted}{ssr}
Definition focuslin7$_{n,m}\ (\ell : \lens{n}m)\ (G : \texttt{endo}_m) : \texttt{morlin}_{n,n}$7 :=
  7$\Lambda T.\ (\texttt{uncurry}\ \ell)_T \circ G_{\dpower{T}{n-m}} \circ (\texttt{curry}\ \ell)_T$7.
Lemma focusN 7$\ell\ G$7 : naturality (focuslin 7$\ell\ G$7).
Definition focus7$_{n,m}\ \ell\ G$7 := 7\textrm{(a morphism packing \texttt{focuslin} $\ell\ G$ and \texttt{focusN} $\ell\ G$)}7.
\end{minted}
In particular, \coqin{focus} and sequential composition satisfy the
following laws, derived from naturality and lens combinatorics.
\begin{minted}{ssr}
Lemma focus_comp : 7$\texttt{focus}\ \ell\ (F \vcomp{} G) = (\texttt{focus}\ \ell\ F) \vcomp{} (\texttt{focus}\ \ell\ G)$7.
Lemma focusM : 7$\texttt{focus}\ (\ell\lcomp\ell')\ G = \texttt{focus}\ \ell\ (\texttt{focus}\ \ell'\ G)$7.
Lemma focusC : 7$\ell \text{ and } \ell' \text{ disjoint} \to$7
  7$(\texttt{focus}\ \ell\ F) \vcomp{} (\texttt{focus}\ \ell'\ G) = (\texttt{focus}\ \ell'\ G) \vcomp{} (\texttt{focus}\ \ell\ F)$7.
Lemma unitary_comp :
  unitary_mor 7$F \to $7 unitary_mor 7$G \to \texttt{unitary\_mor}\ (F \vcomp{} G)$7.
Lemma unitary_focus : unitary_mor 7$G \to \texttt{unitary\_mor}\ (\texttt{focus}\ \ell\ G)$7.
\end{minted}
The law \coqin{focus_comp} states that the sequential composition of morphism
commutes with focusing.
Similarly, \coqin{focusM} states that \coqin{focus} is a morphism for
the composition of lenses.
The law \coqin{focusC} states that the sequential composition of two morphisms focused
through disjoint lenses (i.e. lenses whose codomains are disjoint)
commutes.
The last two lemmas are about unitarity.
Since all circuits can be built from unitary basic gates using sequential
composition and \coqin{focus}, they are sufficient to guarantee
unitarity for all of them.

\section{Proving correctness of circuits}
\label{sec:proving}
Once we have defined a circuit by combining gates through the above
functions, we want to prove its correctness. Usually this involves
proving a relation between the input and the output of the
transformation, which can be expressed as a behavior on computational
basis vectors. In such situations, the following lemmas, from the
\coqin{dpower} module, allow the proof to progress.
\begin{minted}{ssr}
Variables 7$(n\ m : \Nat)\ (\ell : \lens{n}m)\ (T : K\textrm{-module})$7.
Definition dpmerge : 7$I^n \to \dpower{K}m \xrightarrow{\rm linear} \dpower{K}n$7.
Lemma focus_dpbasis : 7$(\texttt{focus}\ \ell\ G)_K\ \ket{v} = \texttt{dpmerge}\ v\ (G_K\ \ket{\texttt{extract}\ v})$7.
Lemma dpmerge_dpbasis : 7$\texttt{dpmerge}\ v\ \ket{v'} = \ket{\texttt{merge}\ v'\ (\extractC\ v)}$7.
Lemma decompose_scaler : 7$\forall(\sigma : \dpower{K}n), \sigma = \sum_{v : \tuple{I}n} \sigma(v) \cdot \ket{v}$7.
\end{minted}
The function \coqin{dpmerge} embeds the result of a quantum gate
applied to a part of the system into the whole system, using the
input computational basis vector for complement; this can be seen as an
asymmetric variant of the \textbf{put} operation.
It is defined using $\texttt{uncurry}$ and \coqin{dpmap}.
It is only introduced and eliminated through the two lemmas following.
The helper law \coqin{focus_dpbasis} allows one to apply
the morphism $G$ to the local part of the basis vector $v$.
The result of this application must then be decomposed
into a linear combination of (local) basis vectors, either by using
the definition of the gate, or by using \coqin{decompose_scaler}.
One can then use linearity to obtain terms of the form
$\sigma(v)\cdot\coqin{dpmerge}\ v\ \ket{v'}$ and merge the local result
into the global quantum state.
Linear algebra computations have good support in \mathcomp, so we do
not need to extend it much.

Extraction and merging only rely on lens-related lemmas, orthogonal to
the linear algebra part.
We have already given some of them in Section~\ref{sec:lens}.
There are many more in the \coqin{lens} module.
While they are instrumental in making proofs
manageable, they do not yet form a complete theory of lenses.
While giving an exhaustive list would not be very useful,
a useful concept we have no mentioned yet is that of \emph{sorted
  lens}, which corresponds to lenses that are monotone functions.
\begin{minted}{ssr}
Definition lens_sorted7$_{n,m} : \lens{n}m \to$7 bool.
Variables 7$(n\ m : \Nat)\ (\ell : \lens{n}m)\ (\ell_1 : \lens{m}p)\ (\ell_2 : \lens{n}p)$7.
Lemma lens_sorted_lensC : lens_sorted 7$\lensC\ell$7.
Lemma lens_sorted_basis : lens_sorted (lens_basis 7$\ell$7).
Lemma lens_sorted_comp :
      lens_sorted 7$\ell \to$7 lens_sorted 7$\ell_1 \to$7 lens_sorted 7$(\ell \lcomp \ell_1)$7.
Lemma eq_lens_sorted :
      7$\ell =_i \ell_2 \to \texttt{lens\_sorted}\ \ell \to \texttt{lens\_sorted}\ \ell_2 \to \ell = \ell_2 :> \texttt{seq}\ \ord{n}$7.
Lemma lens_basis_sortedE : lens_sorted 7$\ell \to \texttt{lens\_basis}\ \ell = \ell$7.
\end{minted}
A variety of lenses are known to be monotone by construction.
In turn, this knowledge allows to prove lens equalities, either
directly, as for \coqin{lens_basis_sortedE}, or at the level of the
list representation, as in \coqin{eq_lens_sorted}, which says that if
two lenses are monotone, and contain the same indices, then their
representations are equal.

Using both linear-algebra and lens-based techniques, we have been able
to prove the correctness of a
number of pure quantum circuits, such as the Shor 9-qubit code or the GHZ
preparation.

\setminted{fontsize=\small}

\section{Concrete examples}
\label{sec:examples}
When working on practical examples we move to more concrete settings.
Namely, we use $\mathbb C$ as the coefficient ring. The indices are
now in $\coqin{I} = \ord{2} = \{0,1\}$ (expect for the qutrit example,
which uses $\coqin{I} = \ord{3} = \{0,1,2\}$).
In this section we use \coq{} notations rather than the mathematical ones
of the previous sections, so as to keep close to the actual code.
Common definitions for qubits can be found in the
\coqin{qexamples_common} module~\cite[\coqin{qexamples_common.v}]{qecc}.

As an example, let us recall the circuit diagram of Shor code (Figure~\ref{fig:shorcode}).
It consists of two smaller components, bit-flip and sign-flip codes
(Figures~\ref{fig:bitflipcode}~and~\ref{fig:signflipcode}), in such a way that
three bit-flip codes are placed in parallel and surrounded by one sign-flip code.
This construction can be expressed straightforwardly as the following
\coq{} code, which can be found in the accompanying
development~\cite[\coqin{qexamples_shor.v}]{qecc}.

\begin{minted}[escapeinside=??]{ssr}
Definition bit_flip_enc : endo?$_3$? :=
  focus [lens 0; 2] cnot ?$\vcomp$? focus [lens 0; 1] cnot.
Definition bit_flip_dec : endo?$_3$? :=
  focus [lens 1; 2; 0] toffoli ?$\vcomp$? bit_flip_enc.
Definition hadamard?$_3$? : endo?$_3$? :=
  focus [lens 2] hadamard ?$\vcomp$? focus [lens 1] hadamard ?$\vcomp$? focus [lens 0] hadamard.
Definition sign_flip_dec := bit_flip_dec ?$\vcomp$? hadamard3.
Definition sign_flip_enc := hadamard3 ?$\vcomp$? bit_flip_enc.
Definition shor_enc : endo?$_9$? :=
  focus [lens 0; 1; 2] bit_flip_enc ?$\vcomp$? focus [lens 3; 4; 5] bit_flip_enc ?$\vcomp$?
  focus [lens 6; 7; 8] bit_flip_enc ?$\vcomp$? focus [lens 0; 3; 6] sign_flip_enc.
Definition shor_dec : endo?$_9$? :=
  focus [lens 0; 3; 6] sign_flip_dec ?$\vcomp$? focus [lens 0; 1; 2] bit_flip_dec ?$\vcomp$?
  focus [lens 3; 4; 5] bit_flip_dec ?$\vcomp$? focus [lens 6; 7; 8] bit_flip_dec.
\end{minted}

We proved that the Shor code is the identity on an error-free channel:
\begin{minted}[fontsize=\small]{ssr}
Theorem shor_code_id :
    (shor_dec 7$\vcomp$7 shor_enc) 7$\ket{i,0,0,0,0,0,0,0,0}$7 = 7$\ket{i,0,0,0,0,0,0,0,0}$7.
\end{minted}

The proof is compositional, relying on lemmas for each subcircuit.
\begin{minted}[fontsize=\small]{ssr}
Lemma cnotE : 7$\texttt{cnot}\ \ket{i, j} = \ket{i, i + j}$7.
Lemma toffoliE7$_{00} : \texttt{toffoli}\ \ket{0,0,i} = \ket{0,0,i}$7.
Lemma hadamardK : 7$\forall T,\ \texttt{involutive hadamard}_T$7.
Lemma bit_flip_enc_ok : 7$\texttt{bit\_flip\_enc}\ \ket{i,j,k} = \ket{i,i+j,i+k}$7.
Lemma bit_flip_toffoli :
  bit_flip_dec 7$\vcomp$7 bit_flip_enc = focus [lens 1;2;0] toffoli.
Lemma sign_flip_toffoli:
  sign_flip_dec 7$\vcomp$7 sign_flip_enc = focus [lens 1;2;0] toffoli.
\end{minted}
The first 3 lemmas describe properties of the matrix representation of
gates, and involve linear algebra computations.
The proof of \coqin{HadamardK} also
involves some real computations about $\sqrt 2$. The remaining 3 lemmas
and the theorem do mostly computations on lenses. In total, there were
about 100 lines of proof.

\begin{figure*}[t]
\begin{minipage}{\textwidth}
\begin{mintedBlock}
\begin{sgoal}
  bit_flip_enc ¦ i, j, k ⟩
\end{sgoal}
\begin{stactic}
rewrite /=.
\end{stactic}
\begin{sgoal}
= focus [lens 0; 2] cnot (focus [lens 0; 1] cnot ¦ i, j, k ⟩)
\end{sgoal}
\begin{stactic}
rewrite focus_dpbasis.
\end{stactic}
\begin{sgoal}
= focus [lens 0; 2] cnot (dpmerge [lens 0; 1] [tuple i; j; k]
                            (cnot ¦ extract [lens 0; 1] [tuple i; j; k]⟩)
\end{sgoal}
\begin{stactic}
9{\color{mediumblue}simpl\_extract}9.
\end{stactic}
\begin{sgoal}
= focus [lens 0; 2] cnot (dpmerge [lens 0; 1] [tuple i; j; k] (cnot ¦ i, j ⟩))
\end{sgoal}
\begin{stactic}
rewrite cnotE.
\end{stactic}
\begin{sgoal}
= focus [lens 0; 2] cnot (dpmerge [lens 0; 1] [tuple i; j; k] ¦ i, i + j ⟩)
\end{sgoal}
\begin{stactic}
rewrite dpmerge_dpbasis.
\end{stactic}
\begin{sgoal}
= focus [lens 0; 2] cnot ¦ merge [lens 0; 1] [tuple i; i + j]
                             (extract (lensC [lens 0; 1]) [tuple i; j; k]) ⟩
\end{sgoal}
\begin{stactic}
9{\color{mediumblue}simpl\_merge}9.
\end{stactic}
\begin{sgoal}
= focus [lens 0; 2] cnot ¦ i, i + j, k ⟩
\end{sgoal}
\end{mintedBlock}
\end{minipage}
\caption{Excerpt of interactive proof of \coqin{bit_flip_enc_ok}}
\label{fig:bitflipenc}
\end{figure*}

\begin{figure*}[t]
\begin{minipage}{\textwidth}
\begin{mintedBlock}
\begin{sgoal}
  (shor_dec 7$\vcomp$7 shor_enc) ¦ i, 0, 0, 0, 0, 0, 0, 0, 0 ⟩
\end{sgoal}
\begin{stactic}
rewrite /=.
\end{stactic}
\begin{sgoal}
= focus [lens 0; 3; 6] sign_flip_dec
    (... (focus [lens 0; 3; 6] sign_flip_enc ¦i,0,0,0,0,0,0,0,0⟩) ...)
\end{sgoal}
\begin{stactic}
transitivity (focus [lens 0; 3; 6] (sign_flip_dec 9$\vcomp$9 sign_flip_enc)
                    ¦i,0,0,0,0,0,0,0,0⟩).
rewrite focus_comp /= focus_dpbasis.
\end{stactic}
\begin{sgoal}
= focus [lens 0; 3; 6] sign_flip_dec (...
    (dpmerge [lens 0; 3; 6] (shor_input i)
       (sign_flip_enc ¦extract [lens 0; 3; 6] (shor_input i)⟩) ...)
\end{sgoal}
\begin{stactic}
set sfe := sign_flip_enc _ _.
\end{stactic}
\begin{sgoal}
= focus [lens 0; 3; 6] sign_flip_dec
    (... (dpmerge [lens 0; 3; 6] (shor_input i) sfe) ...)
\end{sgoal}
\begin{stactic}
rewrite (decompose_scaler sfe) !linear_sum /=.
\end{stactic}
\begin{sgoal}
= \sum_(t : 3.-tuple I) focus [lens 0; 3; 6] sign_flip_dec
       (... (dpmerge [lens 0; 3; 6] (shor_input i) (sfe t *: ¦t⟩)) ...)
\end{sgoal}
\begin{stactic}
apply: eq_bigr => t _.
rewrite !linearZ_LR /= dpmerge_dpbasis.
\end{stactic}
\begin{sgoal}
  sfe t *: focus [lens 0; 3; 6] sign_flip_dec (...
    (¦merge [lens 0; 3; 6] t (extract (lensC [lens 0; 3; 6]) (shor_input i))⟩)
    ...)
= sfe t *: focus [lens 0; 3; 6] sign_flip_dec
    ¦merge [lens 0; 3; 6] t (extract (lensC [lens 0; 3; 6]) (shor_input i))⟩
\end{sgoal}
\begin{stactic}
congr (_ *: focus _ sign_flip_dec _ _).
case: t => -[|a [|b [|c []]]] Ht //=.
9{\color{mediumblue}simpl\_merge}9.
\end{stactic}
\begin{sgoal}
  focus [lens 0; 1; 2] bit_flip_dec (...
    (focus [lens 6; 7; 8] bit_flip_enc ¦ a, 0, 0, b, 0, 0, c, 0, 0 ⟩) ...)
= ¦ a, 0, 0, b, 0, 0, c, 0, 0 ⟩
\end{sgoal}
\end{mintedBlock}
\end{minipage}
\caption{Excerpt of interactive proof of \coqin{shor_code_id}}
\label{fig:shorid}
\end{figure*}

To give a better idea of how the proofs proceed, we show a few steps
of the beginning of \coqin{bit_flip_enc_ok},
in Figure~\ref{fig:bitflipenc}, interspersing
tactics on a gray background between quantum state expressions and equations.
Lines beginning with an ``\coqin{=}'' symbol state that the expression is equal
to the previous one.

Simplifying on line 2 reveals the focused application of the two
\coqin{cnot} gates.
Rewriting with \coqin{focus_dpbasis}, on line 4, applies the
first gate directly to a basis vector.
The helper tactic
\coqin{simpl_extract}~\cite[\coqin{lens_tactics.v}]{qecc}, on line 7,
computes the tuple obtained by \coqin{extract}
(\mathcomp{} is not good at computing in presence of dependent types).
It results here in the vector $\ket{i, j}$, which we can rewrite with
\coqin{cnotE}. As a result, on line 10, \coqin{dpmerge} is
applied to a basis vector, so that we can rewrite it with
\coqin{dpmerge_dpbasis}. Again, on line 14, we use a helper tactic
\coqin{simpl_merge}~\cite[\coqin{lens_tactics.v}]{qecc},
which uses the same code as \coqin{simpl_extract}
to simplify the value of the merge expression. We obtain
$\ket{i,i+j,k}$ as result after the first gate, and can
  proceed similarly with the second gate to reach
$\ket{i,i+j,i+k}$.

As we explained above, our approach cleanly separates computation on
lenses from linear algebra parts. Namely, in the above proof we have
three logical levels:
\coqin{focus_dpbasis} and \coqin{dpmerge_dpbasis}
let one get in and out of a \coqin{focus} application;
\coqin{simpl_extract} and \coqin{simpl_merge} are doing lens
computations; and finally \coqin{cnotE} uses a property
of the specific gate.

The proof of \coqin{shor_code_id} is more
involved as the Hadamard gates introduce superpositions.
It is about 30 lines long.
We show here the first half of the proof in Figure~\ref{fig:shorid}.
The basic idea is to pair the encoders and decoders, and to turn
them into Toffoli gates, which happen to be identities when the extra
inputs are zeros. The first goal is to prove that
\begin{minted}{ssr}
(shor_dec 7$\vcomp$7 shor_enc) 7$\ket{i,0,0,0,0,0,0,0,0}$7 =
focus [lens 0;3;6] (sign_flip_dec 7$\vcomp$7 sign_flip_enc) 7$\ket{i,0,0,0,0,0,0,0,0}$7
\end{minted}
If we expand the compositions on both sides, we see that they both
start by applying \coqin{focus [lens 0;3;6] sign_flip_enc} to the
input (lines 4, 6 and 7).
We can use \coqin{focus_dpasis} to
progress, but due to the Hadamard gates in \coqin{sign_flip_enc}, the
state of the corresponding 3 qubits becomes non-trivial.
However, we can use \coqin{decompose_scaler} on line 14 to see this state as a
sum of unknown computational basis vectors, and progress using linear algebra
lemmas to obtain an equality of sums.
The lemma \coqin{eq_bigr}, from the \coqin{bigop} module of \mathcomp,
allows us to get under the sum, then linear algebra lemmas and
\coqin{dpmerge_dpbasis} let us reach the goal at line 19.
By using congruence, we can get rid of the unknown \coqin{sfe t}
factor, and the enclosing common context.
The \coqin{case} tactic at line 25 decomposes the 3-bit vector \coqin{t}
into 3 separate bits \coqin{a, b, c}, which appear in the simplified
goal at line 27.
We have now reached the bit-flip part of the circuit.
The remainder of the proof consists in using \coqin{focusC} to reorder
the bit-flip encoders and decoders, so that the corresponding ones are
sequentially paired. We can then use \coqin{focus_comp} to produce
applications of \coqin{bit_flip_dec} $\vcomp$ \coqin{bit_flip_enc}, which
can be converted to Toffoli gates by \coqin{bit_flip_toffoli}.
Then we observe that in the input the ancillaries are all zeros, so
that the result of each gate is the identity, which concludes the
first part of the proof. Then we can proceed similarly to prove that
the remaining composition of the sign-flip encoder and decoder is the
identity, which concludes the proof.

\begin{figure}[t]
  \[\ghzfive\]
  \caption{5-qubit GHZ state preparation}
  \label{fig:ghzcircuit}
\end{figure}

Another interesting example is the Greenberger-Horne-Zeilinger (GHZ)
state preparation. It is a generalization of the Bell state, resulting
in a superposition of ${\ket 0}^{\otimes n}$ and ${\ket 1}^{\otimes n}$,
which denote states composed of $n$ zeroes and ones, respectively.
As a circuit, it can be expressed by the composition of one Hadamard
gate followed by $n$ CNOT gates, each one translated by 1 qubit, starting
from the state ${\ket 0}^{\otimes n}$.
The 5-qubit case is shown in Figure~\ref{fig:ghzcircuit}.

\begin{figure*}[t]
\begin{minipage}{\textwidth}
\begin{mintedBlock}
\begin{sgoal}
  lp := lens_pair (succ_neq (n : [n.+1]))
  =======================================
  merge lp [tuple 1; 1]
    (extract (lensC lp) [tuple if i != n.+1 then 1 else 0 | i < n.+2])
  = [tuple 1 | _ < n.+2]
\end{sgoal}
\begin{stactic}
apply eq_from_tnth => i; rewrite [RHS]tnth_mktuple.
case/boolP: (i \in lp) => Hi.
\end{stactic}
\begin{sgoal}
  Hi : i \in lp
  ============================
  tnth (merge lp [tuple 1; 1]
   (extract (lensC lp) [tuple if i0 != n.+1 then 1 else 0 | i0 < n.+2])) i = 1
\end{sgoal}
\begin{stactic}
rewrite tnth_merge -[RHS](tnth_mktuple (fun=>1) (lens_index Hi)).
\end{stactic}
\begin{sgoal}
  tnth [tuple 1; 1] (lens_index Hi) = tnth [tuple 1 | _ < 2] (lens_index Hi)
\end{sgoal}
\begin{stactic}
by congr tnth; 9\color{mediumblue}eq\_lens9.
\end{stactic}
\begin{sgoal}
  Hi : i \notin lp
  ============================
  tnth (merge lp [tuple 1; 1]
   (extract (lensC lp) [tuple if i0 != n.+1 then 1 else 0 | i0 < n.+2])) i = 1
\end{sgoal}
\begin{stactic}
rewrite -mem_lensC in Hi.
rewrite tnth_mergeC tnth_extract tnth_mktuple.
\end{stactic}
\begin{sgoal}
  Hi : i \in lensC lp
  ============================
  (if tnth (lensC lp) (lens_index Hi) != n.+1 then 1 else 0) = 1
\end{sgoal}
\begin{stactic}
rewrite tnth_lens_index ifT //.
\end{stactic}
\begin{sgoal}
  i != n.+1
\end{sgoal}
\begin{stactic}
move: Hi; rewrite mem_lensC !inE; apply contra.
\end{stactic}
\begin{sgoal}
  i == n.+1 -> (i == (n : [n.+2])) || (i == (n.+1 : [n.+2]))
\end{sgoal}
\begin{stactic}
by move/eqP => Hi; apply/orP/or_intror/eqP/val_inj.
\end{stactic}
\end{mintedBlock}
\end{minipage}
\caption{Excerpt of interactive proof of \coqin{ghz_ok}}
\label{fig:ghz}
\end{figure*}

We can write the transformation part as follows in our framework (for
an arbitrary $n$)~\cite[\coqin{qexamples_ghz.v}]{qecc}:
\begin{minted}{ssr}
Lemma succ_neq 7$n\ (i : \ord{n}) : (i : \ord{n+1}) \neq (i+1 : \ord{n+1})$7.
Fixpoint ghz n :=
  match n as n return endo7$_{\tt n.+1}$7 with
  | 0 => hadamard
  | m.+1 => focus (lens_pair (succ_neq (m:[m.+1]))) cnot 7$\vcomp$7
            focus (lensC (lens_single (m.+1:[m.+2]))) (ghz m)
  end.
\end{minted}
The definition works by composing $\coqin{ghz}(m)$, which has type
$\coqin{endo}_n$ (since $n = m+1$), with an extra CNOT gate.
Note that we use dependent types, and the recursion is at a different
type.
The lemma \coqin{succ_neq} is a proof that $i \neq i+1$ in $\ord{n+1}$.
The notation $(i : \ord{m})$ in expressions
denotes that we have a proof that $i \in \ord{m}$; in the actual
code one uses the \coqin{Ordinal} constructor together with some proof
term to build such dependently-typed values.
\coqin{succ_neq} is used by \coqin{lens_pair} to build the lens
$[\coqin{lens}~ m; m+1]$ of type $\lens{m+2}2$.
\coqin{lens_single} builds a singleton lens, so that
\coqin{lensC (lens_single (m.+1:[m.+2]))} is the lens from
$\ord{m+1}$ to $\ord{m+2}$
connecting the inner circuit to the first $m+1$ wires.
We can express the target state and correctness property as follows:
\begin{minted}{ssr}
Definition ghz_state 7$n : \dpower{(\mathbb{C} ^ 1)}{n+1}$7 := 7$\displaystyle\frac{1}{\sqrt{2}} \cdot \left(\ket{0}^{\otimes(n+1)} + \ket{1}^{\otimes(n+1)}\right)$7.
Lemma ghz_ok : ghz 7$n$7 7${\ket{0}}^{\otimes(n+1)}$7 = ghz_state 7$n$7.
\end{minted}
Due to the nesting of lenses, the proof includes a lot of lens
combinatorics, and is about 50 lines long.
We only show the last few lines of the proof in Figure~\ref{fig:ghz},
as they include typical steps. They prove the action of the last CNOT
gate of the circuit when it propagates a $1$ to the last qubit of the
state.
The notation \coqin{[tuple F i | i < 7$n$7]} denotes the $n$-tuple
whose \coqin{i}th element is \coqin{F i}.
Lemma \coqin{eq_from_tnth} on line 6 allows index-wise reasoning.
The \coqin{tnth_mktuple} on the same line extracts the $i$th element
of the tuple comprehension on the right-hand side.
We immediately do a case analysis on whether $i$ is involved in the
last gate. In the first case, we have
$i \in \coqin{lens_pair} (\coqin{succ_neq} (n : \ord{n+1}))$, so we can use
\coqin{tnth_merge} on the left-hand side. On the right-hand side we
use \coqin{tnth_mktuple} backwards, to introduce a 2-tuple.
As a result, we obtain on line 13 a goal on which we can use congruence, and
conclude with \coqin{eq_lens} as both tuples are equal.
The second case, when $i \notin \coqin{lens_pair} (\coqin{succ_neq} (n : \ord{n+1}))$,
is more involved. By using \coqin{mem_lensC} in \coqin{Hi}, we can use
\coqin{tnth_mergeC}, followed by \coqin{tnth_extract} and
\coqin{tnth_mktuple} to reach the goal at line 21. But then the argument
to \coqin{tnth} is precisely that of \coqin{Hi}, so this expression
can be rewritten to $i$ by \coqin{tnth_lens_index}.
From line 25 on it just remains to prove that $i$ cannot be $n+1$, which
is true since it is in the complement of
$\coqin{lens_pair} (\coqin{succ_neq} (n : \ord{n+1}))$.

As we already mentioned in Section~\ref{sec:gate}, our framework is not
limited to qubits, but can accommodate quantum information units with
more states. Here is an example using
qutrits~\cite[\coqin{qexamples_qutrit.v}]{qecc}.
\begin{minted}{ssr}
Definition I : finType := 7$\ord{3}$7.
Definition qnot12 : endo7$_1$7 := dpmor 7$(v : \ord{3} \finmapsto \ket{- v \!\mod 3})$7.
Definition cnot : endo7$_2$7 := dpmor 7$(v : \ord{3}^2 \finmapsto \ket{\tnth{v}0, (\tnth{v}0 + \tnth{v}1 \!\mod 3)})$7.
Definition swap : endo7$_2$7 := dpmor 7$(v : \ord{3}^2 \finmapsto \ket{\tnth{v}1, \tnth{v}0})$7.

Lemma qnot12E : 7$\texttt{qnot12}\ \ket{i} = \ket{-i \mod 3}$7.
Lemma cnotE : 7$\texttt{cnot}\ \ket{i, j} = \ket{i, (i + j \mod 3)}$7.
Lemma swapE : 7$\texttt{swap}\ \ket{i, j} = \ket{i, j}$7.

Lemma swap_cnot_qnot :
  swap =e cnot 7$\vcomp$7 cnot 7$\vcomp$7 focus [lens 1; 0] cnot 7$\vcomp$7 focus [lens 0] qnot12 7$\vcomp$7 cnot.
\end{minted}

Qutrits are represented by normalized vectors in $\mathbb C^3$,
thus we take $\ord 3$ as the indexing type for the basis vectors.
As in the qubit case, the QNOT, CNOT and swap gates are defined and characterized by
their action on indices.
Intuitively, \coqin{cnot} rotates the state of the second
qutrit according to the first one
(rather than flip it), while
\coqin{qnot12} exchanges states 1 and 2, leaving state 0 stationary.

Note that, unlike the qubit QNOT and CNOT, the corresponding qutrit gates
lose their uniqueness because of the added dimension.
There are two other QNOT variants that exchange other combinations of states,
as well as two gates that rotate indices in a single qutrit.
The square of CNOT ($\coqin{cnot} \vcomp \coqin{cnot}$) is different from
\coqin{cnot} but equivalent: they are both different from the identity operator
(if $j \not= 0$) since $i + 2j$, $i + j$ and $i$ are all different,
and each can simulate the other by squaring.

An interesting observation about qutrits is that, contrary to the
qubit case, swapping cannot be done by using only CNOT
gates~\cite{wilmott2008interchangingstatespairqudits}.
Yet there are various ways to define the swap gate using the CNOT
and QNOT gates~\cite[Section IV]{Garcia_Escartin_2013}.

\begin{figure}[t]
\[
\Qcircuit @C=2em @R=1em {
 \ket {\varphi_1} & & \ctrl{1} & \gate{X_d} & \targ     & \ctrl{1} & \ctrl{1} & \qw & \ket {\varphi_2}\\
 \ket {\varphi_2} & & \targ    & \qw        & \ctrl{-1} & \targ    & \targ    & \qw & \ket {\varphi_1}
}
\]
  \caption{Right-hand side of \coqin{swap_cnot_qnot} (borrowing a short name $X_d$ for \coqin{qnot12} from \cite{Garcia_Escartin_2013})}
  \label{fig:swapcnotqnot}
\end{figure}

\begin{figure*}
\begin{minipage}{0.54\textwidth}
\begin{minted}[fontsize=\footnotesize]{ssr}
Lemma swap_cnot_qnot : swap =e ...
Proof.
apply/eq_mor_basis => -[[|i [|j []]] Ht] //.
rewrite /= swapE cnotE focus_dpbasis.
simpl_extract.
rewrite qnot12E dpmerge_dpbasis.
simpl_merge dI.
rewrite focus_dpbasis.
simpl_extract.
rewrite cnotE dpmerge_dpbasis.
simpl_merge dI.
rewrite 2!cnotE addrAC subrr add0r.
have -> : j + (j + (i + j)) = 3 * j + i by ring.
by rewrite (@pchar_Zp 3) // !linE.
Qed.
\end{minted}
\end{minipage}
\begin{minipage}{0.45\textwidth}
\begin{minted}[fontsize=\footnotesize]{ssr}
Lemma addii (i : I) : i + i = 0.
Proof.
by rewrite -(mulr0 i) -(@pchar_Zp 2)//; ring.
Qed.

Lemma swap_cnot : swap =e ...
Proof.
apply/eq_mor_basis => -[[|i [|j []]] Ht] //.
rewrite /= swapE cnotE focus_dpbasis.
simpl_extract.
rewrite cnotE dpmerge_dpbasis.
simpl_merge.
rewrite cnotE addrAC addii add0r.
by rewrite addrCA addii addr0.
Qed.
\end{minted}
\end{minipage}
\caption{Similar proofs of \coqin{swap_cnot_qnot} and \coqin{swap_cnot}}
\label{fig:swap}
\end{figure*}

The last lemma \coqin{swap_cnot_qnot}
(Figure~\ref{fig:swapcnotqnot}),
which provides such a characterization,
can be proved in a way very similar to \coqin{bit_flip_enc_ok}, or to the lemma
\coqin{swap_cnot} that we proved separately in the context of qubits
($\coqin{I} = \ord{2}$).
\begin{minted}{ssr}
Lemma swap_cnot : swap =e cnot 7$\vcomp$7 focus [lens 1; 0] cnot 7$\vcomp$7 cnot.
\end{minted}
We show the respective proofs in Figure~\ref{fig:swap}.
In both cases the first line uses \coqin{eq_mor_basis} to reduce the
problem to basis vectors, and then splits them into individual bits,
in a way similar to the \coqin{shor_code_id} proof. The
rest of the proof mixes lens handling and bit/trit computation.

\section{Parallel composition}
\label{sec:monoid}
In this section, we extend our theory with
noncommutative and commutative monoids of the sequential
and parallel compositions of morphisms.
Thanks to quantum state currying, we have been able to define focusing
and composition of circuits without relying on the Kronecker product.
This also means that parallel composition is not primitive in
this system. Thanks to \coqin{focusC}, morphisms applied through
disjoint lenses do commute, but it is harder to extend this to an
n-ary construct, as done in CoqQ~\cite{CoqQ-popl2023}.
Yet it is possible to define parallel composition using \mathcomp{}
{\em big operators}
by defining a new notion of commuting composition of morphisms.
Note that big operators on monoids require axioms based on propositional
equality, rather than the extensional equality of morphisms,
so in this section (and in the corresponding
development~\cite[\coqin{endo_monoid.v}]{qecc})
we assume functional extensionality and proof irrelevance,
which allows us to use lemma \coqin{morP} of Section~\ref{sec:gate}.

As a first step, we define the noncommutative monoid of morphisms,
using the sequential ({\em vertical}\/ in category-theoretic terminology)
composition as monoid operation and the identity morphism as unit element.
Registering the associativity and unitality laws with Hierarchy Builder~\cite{cohen2020fscd},
allows one to use the corresponding $m$-ary big operator.
\begin{minted}{ssr}
HB.instance Definition _ := Monoid.isLaw.Build 7\rm{}on7 7$\vcomp_{n,n,n}$7 7\rm{}and7 idmor7$_n$7.
Definition compn_mor 7$m\ (F : \ord{m} \to \ndo{n})\ (P : \texttt{pred}\ \ord{n}) :=$7
  \big[7$\vcomp_{n,n,n}$7/idmor7$_n$7]7$_{(i < n ,\ P\ i)}$7 7$F\ i$7.
\end{minted}
By itself, it just allows us to define some circuits in a more compact
way. It will also allow us to connect with the commutative version.

The parallel ({\em horizontal}\/) composition of morphisms is derived from
vertical composition, in the case where the morphisms focused in a circuit
have disjoint supports.
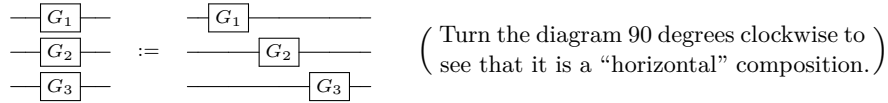
\newcommand\void{|[draw=none]|{}}
\begin{center}
\scriptsize
\(\displaystyle
  \raisebox{0.5cm}{\Qcircuit @C=0.7em @R=.3em {
    & \qw & \gate{G_1} & \qw & \qw\\
    & \qw & \gate{G_2} & \qw & \qw\\
    & \qw & \gate{G_3} & \qw & \qw
  }}
  \quad := \quad
  \raisebox{0.5cm}{\Qcircuit @C=0.5em @R=.3em {
    & \qw & \gate{G_1} & \qw & \qw & \qw & \qw\\
    & \qw & \qw & \gate{G_2} & \qw & \qw & \qw\\
    & \qw & \qw & \qw & \gate{G_3} & \qw & \qw
  }}
  \qquad
  \left(\mbox{
    \begin{minipage}{0.4\linewidth}
      \small
      Turn the diagram 90 degrees clockwise to see that it is a ``horizontal'' composition.
    \end{minipage}
  }\right)
\)
\end{center}
We construct a commutative monoid whose operation is the horizontal
composition, by reifying the notion of focused morphism
(inside an $n$-qubit circuit), using the corresponding lens to express the support.
\begin{minted}{ssr}
Record foc_endo7$_n$7 := 7$\left\{  (m, \ell, e) : \Nat \times \lens{n}m \times \texttt{endo}_m \mid \text{$\ell$ is monotone} \right\}$7.
\end{minted}
The monotonicity of $\ell$
in focused morphisms is demanded for
the canonicity and strictness of their compositions.
The arity $m$ of the morphism is existentially
quantified.

The actual \coq{} definition of \coqin{foc_endo} has four fields \coqin{foc_m}, \coqin{foc_l}, \coqin{foc_e}, and \coqin{foc_s},
the first three corresponding to $m, \ell, e$ above, and the last one being the proof that $\ell$ is monotone.
We define \coqin{mkFendo}, a ``smart constructor'' that factorizes a given lens (\coqin{lens_basis} and \coqin{lens_perm} in Section~\ref{sec:lens})
into its basis (whose monotonicity proof being \coqin{lens_sorted_basis}) and permutation to build a focused morphism.
\begin{minted}{ssr}
Definition 7$\texttt{mkFendo}_{n,m}\ (\ell : \lens{n}{m})\ (G : \texttt{endo}_m)$7 :=
  {| foc_s := lens_sorted_basis 7$\ell$7; foc_e := focus (lens_perm 7$\ell$7) 7$G$7 |}.
\end{minted}
 
Focused morphisms come with both a unit element and an annihilating (zero) element.
\begin{minted}{ssr}
Definition id_fendo  := mkFendo (lens_empty 7$n$7) (idmor I 7$K$7 0).
Definition err_fendo := mkFendo (lens_id 7$n$7) (nullmor 7$n\ n$7).
\end{minted}
The unit element \coqin{id_fendo} has an empty support, and the zero
element \coqin{err_fendo} has a full support.

A focused morphism can be used as an ordinary morphism at arity $n$
by actually focusing the morphism field $e$ along the lens field $\ell$
(field projections \coqin{foc_l} and \coqin{foc_e} are denoted by $.\ell$ and $.e$).
\begin{minted}{ssr}
Definition fendo_mor (7$\Phi$7 : foc_endo) : endo7$_n$7 := focus 7$\Phi.\ell$7 7$\Phi.e$7.
\end{minted}

We can then define commutative composition \coqin{comp_fendo}.
\begin{minted}{ssr}
Definition par_comp7$_{p,q}\ (F : \ndo{p})\ (G : \ndo{q}) : \ndo{p+q} := $7
  (focus lens_left 7$F$7) 7$\vcomp$7 (focus lens_right 7$G$7)
Definition comp_fendo (7$\Phi\ \Psi$7 : foc_endo) :=
  7$\begin{cases}  \texttt{mkFendo}\ (\Phi.\ell\ \texttt{++}\ \Psi.\ell : \lens{n}{\Phi.m+\Psi.m})          \ (\texttt{par\_comp}\ \Phi.e\ \Psi.e) \quad  \hfill \text{if}\ \Phi.\ell\ \text{and}\ \Psi.\ell\ \text{are disjoint} \\   \texttt{err\_fendo} \hfill \text{otherwise} \end{cases}$7
\end{minted}
To make composition commutative, we return the zero element whenever
the lenses of the two morphisms are not disjoint. If they are
disjoint, we return their composition, using the union of the two
lenses. We require lenses to be monotone to guarantee associativity.

Using this definition of commutative composition, we can declare
the commutative monoid structure on focused morphisms and define their
$m$-ary parallel composition. When the lenses are pairwise
disjoint, it coincides with \coqin{compn_mor}.
\begin{minted}{ssr}
HB.instance Definition _ := Monoid.isComLaw.Build 7\rm{}on7 comp_fendo 7\rm{}and7 id_fendo.
Variables 7$(m : \Nat)\ (F : \ord{m} \to \mathtt{foc\_endo})\ (P : \texttt{pred}\ \ord{m})$7.
Definition compn_fendo := \big[comp_fendo/id_fendo]7$_{(i < m ,\ P\ i)}$7 7$F\ i$7.
Hypothesis Hdisj : 7$\forall i, j,\ i \neq j \to (F\ i).\ell\ \textrm{and}\ (F\ j).\ell\ \textrm{are disjoint}$7.
Theorem compn_mor_disjoint :
  compn_mor (fendo_mor 7$\circ$7 F) P = fendo_mor compn_fendo.
\end{minted}

To exemplify the use of this commutative monoid, we proved that the circuit
that consists of $\lfloor n/2\rfloor$ swap gates
that swap the $i$th and $(n-i-1)$th of $n$ qubits
returns a reversed state (Figure~\ref{fig:revcircuit})
\cite[\coqin{qexamples_revcircuit.v}]{qecc}.

\begin{figure}[t]
  \[\revcircuitfive\]
  \caption{5-qubit reversed state circuit}
  \label{fig:revcircuit}
\end{figure}

\begin{minted}{ssr}
Lemma rev_ord_neq7$_n\ (i : \ord{\lfloor n/2\rfloor})$7 : 7$(i : \ord{n}) \neq (n - i - 1 : \ord{n})$7.
Definition rev_circuit 7$n$7 : 7$\ndo{n}$7 :=
  7$\texttt{compn\_mor}\ (i \mapsto \texttt{focus}\ (\texttt{lens\_pair}\ (\texttt{rev\_ord\_neq}\ i))\ \texttt{swap})$7 xpredT.
Lemma rev_circuit_ok : 7$\forall(i : \ord{n}),$7
  proj (lens_single 7$(n-i-1 : \ord{n})$7) (rev_circuit 7$n\ \sigma$7) = proj (lens_single 7$i$7) 7$\sigma$7.
\end{minted}
Here \coqin{rev_ord_neq} produces an inequality in $\ord{n}$, which we can
use to build the required pair lens to apply \coqin{swap}.

\section{Related works}
\label{sec:related}
There are many works that aim at the mechanized verification of
quantum programs~\cite{lewis2022formal}.
Here we only compare with a number of like-minded approaches,
built from first principles, i.e. where the formalization includes a
model of computation based on unitary transformations, which justifies
the proof steps.

Qiskit~\cite{Qiskit} is a framework for writing quantum programs in
Python. While it does not let one write proofs, it has the ability to
turn a circuit into a gate, allowing one to reuse it in other circuits,
so that it has definitional compositionality.

QWIRE~\cite{Qwire-popl2017} and SQIR~\cite{sqir2021popl} define a quantum
programming language and its Hoare logic in
\coq{}\footnotemark, modeling internally
computation with matrices and Kronecker products.
\footnotetext[\thefootnote]{\coq{} was named Coq at the time of these papers.}
QWIRE and SQIR differ in their handling of
variables: in QWIRE they are abstract, handled through higher-order
abstract syntax, but in SQIR, which was originally intended as an
intermediate language for the compilation of QWIRE, they are concrete
natural numbers, denoting indices of qubits. The authors note in their
introduction~\cite{sqir2021popl} that ``[abstract variables]
necessitate a map from variables to indices, which we find confounds
proof automation''.
They go on remarking that having a distinct semantics for pure
quantum computation, rather than relying only on the density matrices
needed for hybrid computations, considerably simplifies proofs; this
justifies our choice of treating specifically the pure case.
While QWIRE satisfies definitional compositionality, this is not the
case for SQIR, as circuits using fixed indices cannot be directly
reused.
We have not proved enough programs to provide a meaningful
comparison, yet it is noteworthy that our proof of GHZ, which uses
virtually no automation, is about half the size of the proof in
SQIR~\cite{sqir2021popl}. The main difference is that we are
able to solve combinatorics at the level of lenses, while they
have to work all along with a symbolic representation of matrices,
that is a linear combination of matrix units (Dirac's notation), to avoid
working directly on huge matrices.

VyZX~\cite{lehmann2022vyzx,lehmann2023vyzx} formalizes the ZX-calculus
in \coq{}\footnotemark[\thefootnote], on top of SQIR. Its goal is to
prove graph-rewriting rules, and
ultimately to build a verified optimizer for the ZX-calculus. However, as
they state themselves, the
graphical nature of the calculus appears to be a major difficulty, and
only restricted forms of the rules are proved at this point. Since the
ZX-calculus itself enjoys compositionality, albeit at the graph level,
this is a promising line of work. It would be interesting to see if
our approach can make proving such graph-rewriting rules easier.
As preliminary experiment, we have proved the triangular identity
involving a cup and a cap, by defining an asymmetric version of focusing.
More generally, finding a nice way to compose graphs is
essential, and concepts such as lenses could have a role there.

CoqQ\footnotemark[\thefootnote]~\cite{CoqQ-popl2023} builds a
formalized theory of Hilbert spaces
and n-ary tensor products on top of \mathcomp{}, adding support for the
so-called {\em labelled Dirac notation}. Again they define a
Hoare logic for quantum programs, and are able to handle both pure and
hybrid computations. While the labelled Dirac notation allows handling
commutation comfortably, it does not qualify as compositional,
since it is based on a fixed set of labels, i.e. one cannot mix
programs if they do not use the same set of labels.

Unruh developed a quantum Hoare logic and formalized it in Isabelle,
using a concept of {\em register}~\cite{Unruh-planqc21} for which
he defines a theory, including operations such as taking the
complement of a register.
His registers in some meaning generalize our \coqin{focus} function,
as they allow focusing between arbitrary types rather than just sets
of qubits.
Since one can compose registers, his approach
is compositional, for both definitions and proofs, and the abstraction
overhead is avoided through automation.
However, while each application of \coqin{focus} to a lens
can be seen as a register, he has not separated out a concrete
combinatorics based on finite objects similar to our notion of lens.

In a slightly different direction, Qbricks~\cite{Chareton2021} uses
the framework of {\em path-sums} to allow the automatic proof of pure
quantum computations. The notion of path is more expressive than that
of computational basis state, and allows one to represent many unitary
transformations as maps from path to path, making calculations
easier. It would be interesting to see whether it is possible to use
them in our framework.

Most approaches above support not only pure quantum computation but
also hybrid quantum-classical computation. While we have concentrated
here on pure computation, we have already extended our approach to the
density-matrix interpretation required to support hybrid computations,
and verified that it commutes with focusing.
Practical applications are left to future work.

Note also that, while some of the above works use dependent types to
represent matrix sizes for instance, they all rely on ways to hide or
forget this information as a workaround. On the other hand, our
use of dependent types is strict, only relying on statically proved
cast operators to adjust types where needed, yet it is lightweight
enough for practical use.

Some other aspects of our approach can be related to programming
language theory. Wadler defined parametricity in terms of
naturality, showing its flexibility in reasoning on
programs~\cite{wadler1989fpca}. Our definition of natural morphisms
differs from Wadler's only in that we restrict the domain of discourse
from all types to those representing modules on a ring (or more precisely,
the category of modules).
Another instance is the way we shift indices during currying. It is
reminiscent of De Bruijn indices, and our \coqin{merge} operation
shifts indices in the precise same way as the record concatenation
defined in the label-selective $\lambda$-calculus~\cite{garrigue94a}.
This suggests that our currying of quantum states is actually similar
to the currying occurring in that calculus.

From a category-theoretic viewpoint,
the quantum focusing we defined is closely related to the
generalization of lenses to
\emph{optics}~\cite[Section 2.1]{riley2018arxiv},
which is motivated by applications
to functional programming~\cite{clarke2024compositionality}.
Optics are defined generally for symmetric monoidal categories,
and presented in terms of the category-theoretic tool of
coends~\cite{maclane1998,roman2021act} to cleanly describe
structure among objects.
Our account of quantum lenses and focusing is
specialized to the symmetric monoidal category of modules,
concentrating on concrete representation of objects
rather than trying to show the abstract structure.

\section{Conclusion}
We have been able to build a compositional model of pure quantum
computation in \coq, on top of the \mathcomp{} library, by using
finite functions, lenses, and focusing.
We have applied the development to prove the correctness of
several quantum circuits.
An interesting remark is that, while we started from the traditional
view of seeing quantum states as tensor products, our implementation
does not rely on the Kronecker product for composing
transformations. Since the Kronecker product of matrices  can be
cumbersome to work with, this is a potential advantage of this
approach.

Many avenues are open for future work.
First we need to finish the proof of the Shor code, this time for
erroneous channels; paper proofs are simple enough but the devil is
in the details.
Next, building on our experience, we would like to formalize and abstract
the algebraic theory of lenses. Currently we rely on a large set of
lemmas developed over more than a year, without knowing their
interdependencies; such a theory would have both theoretical and
practical implications.
Third, we are interested in the category-theoretic aspects of this approach,
and would like to relate our approach to \coqin{focus} to the theory of optics,
explaining both the relation between the combinatorics of a lens and its action,
and the structural properties of focusing.

\section*{Statements and declarations}

\noindent
Competing interests:
\begin{itemize}
\item Funding: This research was partially supported by JSPS KAKENHI grant No. 	JP22H00520 and MEXT Q-LEAP grant No. JPMXS0120319794.
\item Employment: The authors are employed by Nagoya University
\item Financial interests: None
\item Non-financial interests: Q-LEAP project, AIST
\end{itemize}

\noindent
Other declarations:
\begin{itemize}
\item Ethics approval and consent to participate: Not applicable
\item Consent for publication: Not applicable
\item Data availability: Not applicable
\item Materials availability: Not applicable
\item Code availability: \url{https://github.com/t6s/qecc/tree/jar2026}
\item Author contribution: Not applicable
\end{itemize}





\bibliography{ttquantum}


\end{document}